\documentclass[preprint,superscriptaddress,showpacs,tightenlines,showkeys]{revtex4}
\usepackage{amssymb}
\usepackage{amsmath}
\usepackage{graphicx,bm}

 
\input{tcilatex}

\begin{document}

\title{Theory of oscillations in the STM conductance resulting from
subsurface defects\\
(Review Article) }
\author{Ye.S. Avotina}
\affiliation{B.I. Verkin Institute for Low Temperature Physics and Engineering, National
Academy of Sciences of Ukraine, 47, Lenin Ave., 61103, Kharkov, Ukraine.}
\author{Yu.A. Kolesnichenko}
\affiliation{B.I. Verkin Institute for Low Temperature Physics and Engineering, National
Academy of Sciences of Ukraine, 47, Lenin Ave., 61103, Kharkov, Ukraine.}
\author{J.M. van Ruitenbeek}
\affiliation{Kamerlingh Onnes Laboratorium, Universiteit Leiden, Postbus 9504, 2300
Leiden, The Netherlands.}

\begin{abstract}
In this review we present recent theoretical results concerning
investigations of single subsurface defects by means of a scanning tunneling
microscope (STM). These investigations are based on the effect of quantum
interference between the electron partial waves that are directly
transmitted through the contact and the partial waves scattered by the
defect. In particular, we have shown the possibility imaging the defect
position below a metal surface by means of STM. Different types of
subsurface defects have been discussed: point-like magnetic and non-magnetic
defects, magnetic clusters in a nonmagnetic host metal, and non-magnetic
defects in a $s$-wave superconductor. The effect of Fermi surface anisotropy
has been analyzed. Also, results of investigations of the effect of a strong
magnetic field to the STM conductance of a tunnel point contact in the
presence of a single defect has been presented.
\end{abstract}

\keywords{STM, electron tunneling, subsurface defect, conductance, Friedel
oscillations}
\keywords{STM, electron tunneling, subsurface defect, conductance, Friedel
oscillations}
\pacs{%
61.72.J
Point
defects
and
the
defect
clusters,
73.40.Cg
Contact
resistance,
contact
potential,
73.63.Rt
Nanoscale
contact,
74.50.+v
Tunneling
phenomena:
single
particle
tunneling and
STM,
73.23.-b
Electronic
transport
in
mesoscopic
systems,72.10.Fk
Scattering
by
point
defects,
dislocations,
surfaces, and
other
imperfections
(including Kondo
effect)%
}
\maketitle
\tableofcontents

\section{Introduction}

About three decades following its invention \cite{Binnig}, scanning
tunnelling microscopy (STM) has proved to be a superbly valuable tool for
investigating surfaces on the atomic scale. Along with a mapping of the
conductor's surface, the STM enables observing many phenomena, among which
electron scattering by single surface defects (impurity atoms, adatoms, or
step edges). There are hundreds of papers that are devoted to investigations
of surface defects by STM. In this paper we do not aim to review all of them
and confine ourselves to briefly mentioning the main directions of
researches in this field. Our attention will be mainly focused on
interference effects in STM conductance caused by defects sitting \emph{below%
} the surface.

Electron scattering by defects leads to quantum-interference patterns in the
local electron density of states around the defects (Friedel oscillations 
\cite{Friedel}). For more than thirty years Friedel oscillations have
remained a theoretical prediction that could be seen only in theory
textbooks \cite{Ziman}. The appearance of the STM has enabled the
visualization of these oscillations, which manifest themselves as
oscillations of the differential tunneling conductance, $G=dI/dV$, around
defects on the surface.

First standing wavelike patterns in the STM conductance in the vicinity of
defects were observed by Crommie \textit{et al.} \cite{Crommie} on a Cu(111)
surface and by Hasegava \textit{et al.} \cite{Hasegava} on a Au(111)
surface. At the (111) surface of the noble metals Cu, Ag, and Au the
electrons of the surface states form a quasi-two-dimensional nearly-free
electron gas having an isotropic dispersion law \cite{Burgi}. When scattered
from step edges or adatoms the surface states form standing waves which
result in an oscillatory dependence of the tunneling conductance measured as
a function of the distance between the STM tip and the defect, $r_{0}$. The
period of the conductance oscillations $\Delta r_{0}=2\pi /2k_{\mathrm{F}%
}^{2D}$ is set by twice the Fermi wave vector, $2k_{\mathrm{F}}^{2D}$ ($%
\mathbf{k}_{\mathrm{F}}^{2D}$ is a two-dimensional vector in the plane of
the surface).

The circular 2D Fermi contour of the electrons at the (111) surface of noble
metals results from the fact that the layer of surface atoms actually
corresponds to one of the close-packed stackings on which the face-centered
cubic structure is based. Generally, for less closely packed surfaces and
conductors having a complicated crystallographic structure a 2D Fermi
contour is anisotropic, i.e. the absolute value of the vector $\mathbf{k}_{%
\mathrm{F}}^{2D}$ depends on its direction. The Fourier transform (FT) of
the standing wave pattern provides an image of the Fermi contour.
Anisotropic Friedel-like oscillations have been observed by FT-STM on
Cu(110) surfaces \cite{Petersen2}, Be \cite{Sprunger}, and ErSi$_{2}$ \cite%
{Vonau}. Particularly, in Ref. \cite{Petersen2} the contour related to the
'neck' of the bulk Fermi surface for Cu (110) surface has been imaged.

Magnetic adatoms on non-magnetic host metal surfaces are of special interest
as they produce a characteristic many-body resonance structure in the
differential conductance near zero voltage bias attributed to the Kondo
effect \cite{chen99,madhavan98,li98,knorr02}. The shape of the resonance in
the differential conductance is usually asymmetric and is described by a
Fano line shape \cite{fano61,plihal01,wahl04}. The surface electron waves
carry information on the magnetic impurity and by focussing the waves it has
been possible to create a mirage image of the impurity \cite{manoharan00}
(for review, see \cite{corral}). The interesting phenomenon of an \emph{%
orbital} Kondo resonance was observed by STM in Ref.\cite{Koles}. It was
found that STM images of the Cr(001) surface show cross-like depressions
centered around the impurities corresponding to the orbital symmetry of two
degenerate $d_{xz},$ $d_{yz}$ surface states \cite{Koles}.

The investigation of defects near the surface of unconventional
superconductors by STM is a way to determine the symmetry of the order
parameter. The effect of single Zn defects on the superconductivity in high-T%
$_{\mathrm{c}}$ superconductors was investigated in Ref.~\cite{Zn}, and the
manifestation of d-wave pairing symmetry was observed in the quasibound
state near the defect. In Ref.~\cite{MagImp} a bound state near a magnetic
Mn adatom on the surface of superconducting Nb was observed by STM.

An effective way to enhance the STM sensitivity to such oscillation effects
is to use a superconducting tip \cite{Pan}. In Ref.~\cite{Xu} it was
demonstrated that the amplitude of conductance oscillations is significantly
enhanced when a superconducting tip is used, and when the applied bias is
close to the gap energy of the superconductor.

The applicability of STM can be extended to the study of magnetic objects on
the surface of a conductor when a magnetic material is used for the STM tip
such that the electric current is spin polarized (SP) (for review of SP-STM
see \cite{Bode}). For example, the precession of a magnetic moment of
clusters of organic molecules on a surface gives rise to a time modulation
of the SP-STM current, from which the $g$ - factor can be found \cite%
{Manoharan,Durkan}. The possibility to probe magnetic properties of
nanostructures buried beneath a metallic surface by means of local probe
techniques is discussed in Ref.\cite{Brovko}. It has been shown that those
properties can be deduced from the spin-resolved local density of states
above the surface \cite{Brovko}.

STM spectroscopy also provides access to information on the structure of the
metal \emph{below} the surface in both semiconductors and metals. Crampin 
\cite{Crampin} proposed to utilize the surface states for imaging subsurface
impurities. However, the exponential decay of the wave function amplitude
into the bulk limits the effective range to the topmost layers only and bulk
states form a good alternative for detecting defect positions. The principle
of imaging subsurface defects is based on the influence on the conductance
caused by quantum interference of electron waves that are scattered by
defects and reflected back by the contact. This effect was explored for
investigating subsurface Ar bubbles submerged in Al \cite{Schmid} and Cu 
\cite{Kurnosikov}, and Si(111) step edges buried under a thin film of Pb 
\cite{Hongbin}. In these experiments, bulk electrons are found to be
confined in a vertical quantum well between the surface and the top plane of
the object of interest. The observation of interference patterns due to
electron scattering by Co impurities in the interior of a Cu sample was
reported Refs. \cite{Quaas,Weismann}.

Reviews of STM theory can be found in Refs.~ \cite{Hofer,Blanco}. The papers
listed in \cite{Hofer,Blanco}, in which the conductance of a tunnel contact
of small size has been analyzed theoretically, must be complemented by
reference to the fundamental paper of Kulik, Mitsai and Omelyanchouk \cite%
{KMO} published in 1974. In this paper the authors obtained, on the basis of
rigorous quantum-mechanical considerations, an analytical formula for the
conductance of a junction between two metal half-spaces separated by an
inhomogeneous tunnel barrier of low transparency. Their result is valid for
arbitrary values of the applied bias and for arbitrary dependence of the
tunnelling probability on the coordinates in the plane of the interface
between the metals. As a special case, the general formula for the contact
resistance can be applied to an inhomogeneous tunnel contacts having a
characteristic diameter smaller than electron wave length, which is suitable
to describe STM conductance. Recently, electron tunnelling through a
randomly inhomogeneous barrier of arbitrary amplitude has been analyzed
theoretically in Refs.~\cite{Bezak1,Bezak2}.

The theoretical descriptions of STM conductance oscillations due to electron
scattering by single defects in the majority of papers is based on the
assumption that the tunnelling conductance measured by the STM tip is
proportional to the local density of states (LDOS) $\nu (\mathbf{r})$ of the
sample (see, for example, \cite{Crampin,corral,Balatsky,Stepanyuk}) as for a
planar tunnel junction \cite{Wolf}. For the scattering of electron surface
states this assumption is quite reasonable, but for electron scattering in
the bulk of the sample it can not be used. The LDOS in the vicinity of
defects in the bulk is critically modified by electron reflections off the
surface of the conductor, at $\mathbf{r\in \Sigma }$, and differs from
Friedel oscillations of the LDOS in an infinite conductor with a single
scatterer \cite{Ziman}. In the limit of zero tunnelling probability we have $%
\nu (\mathbf{r\in \Sigma })=0.$ Further, the conductance oscillations are
formed only by "tagged" electrons, which tunnel through the contact and are
scattered back by the defect, while a "halo" of Friedel oscillations around
the defect is due to all scattered electrons. In general, there are no other
periods in the interference effects but the period of Friedel oscillations $%
\Delta r_{0}=2\pi /2k_{\mathrm{F}}$ ($\mathbf{k}_{\mathrm{F}}$ is a Fermi
wave vector) and the analysis in Ref.~\cite{Weismann} of the experimental
data in terms of a bulk LDOS seems to be qualitatively correct \cite%
{Kobayashi}. However, the calculation of amplitudes and phases of the
conductance oscillations, which contain additional information on the
interaction of the charge carriers with the defect, requires the solution of
the scattering problem of the influence of subsurface defects on the
conductance of a small tunnel contact.

In this paper we review a series of publications in which the theory of the
electronic transport though a tunnel point-contact in the presence of a
single defect below metal surface was developed. The organization of this
paper is as follows. The model of the tunnel contact and the basic equations
that describe the effect of subsurface defects on the STM conductance are
presented in Sec.~II. The solution of the Schr\"{o}dinger equation for
elections that tunnel through the contact and are scattered by the defect is
given. In Sec.~III a method to determine the defect positions below a metal
surface is formulated on the basis of an investigation of the nonlinear
conductance of the contact. A signature of the Fermi surface anisotropy in
STM conductance in the presence of subsurface defects is discussed in
Sec.~IV. In Sec.~V we present the results of investigations of the effect of
a subsurface magnetic defect on the tunnel current, including the signature
of a Kondo impurity and that of a magnetic cluster having an unscreened
magnetic moment. In Sec.~VI it is shown that a strong magnetic field leads
to specific magneto-quantum oscillation periods which depend on the distance
between the contact and the defect. The possibilities of studying the
interference of quasiparticles in a superconductor is analyzed in Sec.~VII.
In Sec.~VIII we conclude by discussing the possibilities for exploiting
these theoretical results for sub-surface imaging along with experimental
investigations of physical characteristics of subsurface defects.

\section{Quantum interference of scattered electron waves in the vicinity of
a point contact}

\subsection{Model of a STM contact, and the Schr\"{o}dinger equation for the
system}

\begin{figure}[tbp]
\includegraphics[width=8cm]{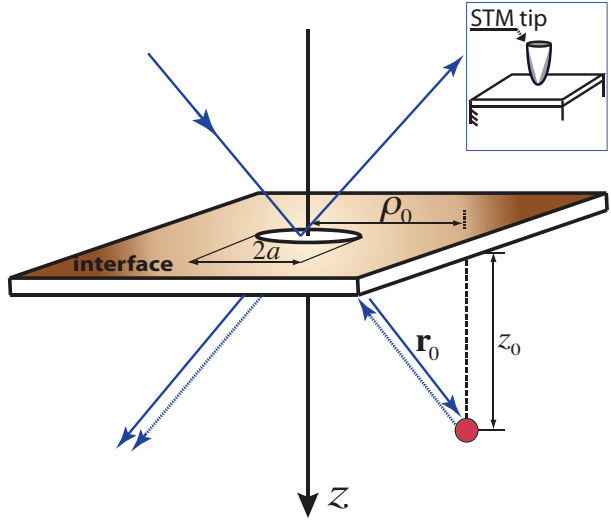}
\caption{Model of the tunnel point contact as an orifice in an interface
that is nontransparent for electrons except for a circular hole, where
tunnelling is allowed. Trajectories are shown schematically for electrons
that are reflected from or transmitted through the contact and then
scattered back by a defect.}
\label{Contact}
\end{figure}

As a model for the STM experiments we choose an inhomogeneous tunnel contact
between two metal half-spaces separated by an infinitely thin interface. The
potential barrier in the plane of the interface, at $z=0$, is taken to be
described by a delta function \cite{KMO}, 
\begin{equation}
U(\mathbf{r})=U_{0}f(\mathbf{\rho })\delta (z),  \label{U}
\end{equation}%
where $\mathbf{\rho }$ is the radius vector in the plane of the interface,
perpendicular to the $z$ axis. The function $f(\mathbf{\rho })\rightarrow
\infty $ at all points of the plane $z=0$ except for a small region defining
the contact, having a characteristic radius $a$, at which $f(\mathbf{\rho })$
is of order 1. As an example, a suitable model for the function $f(\mathbf{%
\rho })$ for the "STM tip" is the Gaussian function $f(\mathbf{\rho })=\exp
(\rho ^{2}/a^{2})$ with small $a.$ Another useful model of the junction is
an orifice of radius $a$ for which $f(\rho )=1$ for $\rho \leq a$ in the
plane of the contact (Fig. \ref{Contact}).

Of course, such a model describes only the qualitative features of the
conductance of an STM contact, and does not contain such parameters as the
tip radius, or the distance between the STM tip and the sample as is
represented, for example, in the model by Tersoff and Hamann \cite{Tersoff}.
In principle these properties of the system may be included in the model as
parameters of the function $f( \mathbf{\rho }).$ The advantage of the model
by Kulik \textit{et al.} \cite{KMO} is the possibility of finding exact
analytical solutions of the Schr\"{o}dinger equation in the limit $%
U_{0}\rightarrow \infty $. The equations are considerably simplified in the
case of a small contact $a\rightarrow 0.$ The wave functions obtained in the
framework of the model barrier (\ref{U}) properly describe the spreading of
electron waves into the bulk metal from a small region on its surface. A
numerical value for the STM conductance plays the role of a scale factor for
the conductance oscillations, and for the further considerations below it is
of less importance.

A defect in the vicinity of the interface can be described by the potential 
\begin{equation}
D(\mathbf{r})=gD_{0}(\left\vert \mathbf{r-r}_{0}\right\vert ),  \label{D(r)}
\end{equation}%
where $g$ is the constant of the electron interaction with the defect, and $%
D_{0}(\left\vert \mathbf{r-r}_{0}\right\vert )$ is a spherically symmetric
function localized within a region of characteristic radius $r_{D}$ centered
at the point $\mathbf{r}=\mathbf{r}_{0}$, which satisfies the normalization
condition 
\begin{equation}
4\pi \int dr^{\prime }r^{\prime 2}D_{0}(r^{\prime })=1.  \label{intD=1}
\end{equation}%
The electron wave function $\psi \left( \mathbf{r}\right) $ in a metal with
a dispersion relation $\varepsilon \left( \mathbf{k}\right) $ must be found
from the Schr\"{o}dinger equation \cite{LAK} 
\begin{equation}
\left[ {\varepsilon }\left( \widehat{\mathbf{k}}-\frac{e}{c\hbar }\mathbf{A}%
\right) +\sigma g_{e}\mu _{\mathrm{B}}H+eV(\mathbf{r})+D(\mathbf{r})+U(%
\mathbf{r})\right] {\psi }={\varepsilon \psi .}  \label{Schrod}
\end{equation}%
Here $\widehat{\mathbf{k}}=-i\nabla ,$ $\mathbf{A}(\mathbf{r})$ is the
vector-potential of the stationary magnetic field $\mathbf{H}$, and $V(%
\mathbf{r})$ is the applied electrical potential, $\sigma =\pm 1$
corresponds to different spin directions, $\mu _{\mathrm{B}}=e\hbar /2m_{0}c$
is the Bohr magneton, where $m_{0}$ is the free electron mass, and $g_{e}$
is the electron $g$-factor. The function $\psi \left( \mathbf{r}\right) $
satisfies at $z=0$ the following boundary condition for continuity of the
wave function 
\begin{equation}
\psi \left( \mathbf{\rho },+0\right) =\psi \left( \mathbf{\rho },-0\right) ,
\label{equal}
\end{equation}%
and the condition, which for a $\delta -$function barrier is obtained by the
integration of the Schr\"{o}dinger equation (\ref{Schrod}) over an
infinitesimal interval near the point $z=0$ 
\begin{equation}
\int\limits_{-0}^{+0}dz{\varepsilon }\left( \widehat{\mathbf{k}}-\frac{e}{%
c\hbar }\mathbf{A}\right) \psi \left( \mathbf{\rho },z\right) =-U_{0}f(%
\mathbf{\rho })\psi \left( \mathbf{\rho },0\right) .  \label{jump}
\end{equation}%
\qquad In this section below we consider a solution of Schr\"{o}dinger
equation (\ref{Schrod}) for a free electron model with an electron effective
mass $m^{\ast }$ and a dispersion relation $\varepsilon \left( \mathbf{k}%
\right) =\hbar ^{2}\mathbf{k}^{2}/2m^{\ast }$ in the absence of external
fields ($H=0,V=0$ ). In this case the condition (\ref{jump}) reduces to the
well-known condition for the jump of the derivative of the wave function 
\begin{equation}
\psi _{z}^{\prime }\left( \mathbf{\rho },+0\right) -\psi _{z}^{^{\prime
}}\left( -\mathbf{\rho },0\right) =\frac{2m^{\ast }U_{0}}{\hbar ^{2}}f\left( 
\mathbf{\rho }\right) \psi \left( \mathbf{\rho },0\right) .  \label{jump1}
\end{equation}
The effects of applied voltage, Fermi surface anisotropy and magnetic field
are discussed in next sections.

\subsection{Wave function due to an inhomogeneous tunnel barrier}

Here we follow the procedure for the finding the electron wave function in
the limit $U_{0}\rightarrow \infty$ that was proposed in Ref.~\cite{KMO}. To
first approximation in the small parameter $1/U_{0}$ the wave function $\psi
\left( \mathbf{r}\right) $ can be written as: 
\begin{equation}
\psi \left( \mathbf{r}\right) =\psi _{0}\left( \mathbf{r}\right) +\varphi
_{0}\left( \mathbf{r}\right) ,  \label{_psi}
\end{equation}%
where $\varphi _{0}$ is of order $1/U_{0}.$ This latter part of the wave
function (\ref{_psi}) describes the electron tunnelling through the barrier
and determines the electrical current. The first term in the Eq.(\ref{_psi})
is the solution of the Schr\"{o}dinger equation for the metallic half-spaces
without the contact \ 
\begin{equation}
\psi _{0}\left( \mathbf{r}\right) =e^{i\mathbf{\kappa \rho }}\left(
e^{ik_{z}\left\vert z\right\vert }-e^{-ik_{z}\left\vert z\right\vert
}\right) ,  \label{psi00}
\end{equation}%
where $\mathbf{\kappa }$ and $k_{z}$ are the components of the wave vector $%
\mathbf{k}$ parallel and perpendicular to the interface, respectively. The
expression (\ref{psi00}) satisfies the boundary condition $\psi _{0}\left( 
\mathbf{\rho },0\right) =0$ at the interface.

Substituting the wave function (\ref{_psi}) into the boundary conditions (%
\ref{equal})and (\ref{jump}) one must match terms of the same order in $%
1/U_{0}.$ As a result the conditions (\ref{equal}), (\ref{jump}) are reduced
to \cite{KMO} 
\begin{equation}
\varphi _{0}\left( \mathbf{\rho },+0\right) =\varphi _{0}\left( \mathbf{\rho 
},-0\right) ,  \label{equal2}
\end{equation}%
\begin{equation}
t\left( k_{z}\right) e^{i\mathbf{\kappa \rho }}=f\left( \mathbf{\rho }%
\right) \varphi _{0}\left( \mathbf{\rho },0\right) .  \label{jump2}
\end{equation}%
where%
\begin{equation}
t\left( k_{z}\right) =\hbar ^{2}k_{z}/im^{\ast }U_{0};\qquad |t|\ll 1,
\label{t(k)}
\end{equation}%
is the amplitude of the electron wave function passing through the
homogeneous barrier. Developing the function $\varphi _{0}\left( \mathbf{%
\rho },z\right) $\ as a Fourier integral in the coordinate $\mathbf{\rho }$,
and using the Eq.~(\ref{jump2}), we find \cite{KMO} 
\begin{equation}
\varphi _{0}\left( \mathbf{\rho },z\gtrless 0\right) =\frac{t\left(
k_{z}\right) }{\left( 2\pi \right) ^{2}}\int\limits_{-\infty }^{\infty }d%
\mathbf{\kappa }^{\prime }e^{i\mathbf{\kappa }^{\prime }\mathbf{\rho +}%
ik_{z}^{\prime }\left\vert z\right\vert }\int\limits_{-\infty }^{\infty }d%
\mathbf{\rho }^{\prime }\frac{e^{i\left( \mathbf{\kappa -\kappa }^{\prime
}\right) \mathbf{\rho }^{\prime }}}{f\left( \mathbf{\rho }^{\prime }\right) }%
,  \label{phi0}
\end{equation}%
where $k_{z}^{\prime }=\sqrt{k^{2}-\kappa ^{\prime 2}}.$ For a homogeneous $%
\delta $-function barrier, $f(\rho)=1$, Eq.~(\ref{phi0}) transforms into a
transmitted plane wave having an amplitude $t$.

The characteristic radius of the region on the surface through which
electrons tunnel from the STM tip into the sample is of atomic size, $%
a\simeq 0.1\text{\AA } $, while the Fermi wave vector $k_{\mathrm{F}}\simeq 1
$\AA $^{-1}.$ By using the condition $k_{\mathrm{F}}${}$a_{1,2}\ll 1$ after
integrating over $\mathbf{\kappa }^{\prime }$ and $\mathbf{\rho }^{\prime }$
in Eq.(\ref{phi0}) we find \cite{Avotina08mi} 
\begin{equation}
\varphi _{0}\left( \mathbf{r}\right) =t\left( k_{z}\right) \frac{i\left(
ka\right) ^{2}z}{2r}h_{1}^{\left( 1\right) }\left( kr\right) .  \label{phi01}
\end{equation}%
The incident plane wave is transformed into a spherical p-wave $%
h_{1}^{\left( 1\right) }\left( kr\right) $ (\ref{phi01}) after scattering by
the point contact. In Eq.~(\ref{phi01}), and below, $h_{l}^{\left( 1\right)
}\left( x\right) $ are the spherical Hankel functions. Note that the wave
function $\varphi _{0}(\mathbf{r})$ (\ref{phi01}) is zero in all points on
the surface $z=0$, except the point $r=0$ (at the contact) where it
diverges. This divergence is the result of taking the limit $a\rightarrow 0$
in the integral expressions for $\varphi _{0}(\mathbf{r})$ (\ref{phi0}).
Yet, Eq.~(\ref{phi01}) gives a finite value for the total charge current
through the contact as obtained by integration over a half-sphere of radius $%
r$ with its center in the point $\mathbf{r}=0$ for $r\rightarrow 0$. 

\subsection{Electron scattering by a single defect in the vicinity of a
tunnel point contact}

As a result of current spreading only a small region near the point contact
noticeably influences the conductance. For high purity samples only a few
defects will be found in this region. At low temperatures the distance
between the contact and the nearest defect, $r_{0}$, is smaller than the
electron mean free path due to electron-phonon scattering and the electrons
are elastically scattered by the single defect only. The wave function of
transmitted electrons, $\varphi \left( \mathbf{r}\right) $, which takes into
account the scattering by the defect, can be expressed in terms of the
retarded Green function $G_{0}^{+}\left( \mathbf{r,r}^{\prime };\varepsilon
\right) $ of the homogeneous equation (\ref{Schrod}) at $D=0$, in absence of
impurity scattering. To first approximation in the transmission amplitude $t$
(\ref{t(k)}) the integral equation for $\varphi \left( \mathbf{r}\right) $
is given by 
\begin{equation}
\varphi {\left( \mathbf{r}\right) =}\varphi _{0}{\left( \mathbf{r}\right) +}g%
{\int d\mathbf{r}^{\prime }D\left( \left\vert \mathbf{r}^{\prime }-\mathbf{r}%
_{0}\right\vert \right) G_{0}^{+}\left( \mathbf{r,r}^{\prime };\varepsilon
\right) \varphi \left( \mathbf{r}^{\prime }\right) ,}  \label{phi}
\end{equation}%
where 
\begin{equation}
G_{0}^{+}\left( \mathbf{r,r}^{\prime };\varepsilon \right) =-\frac{ikm^{\ast
}}{2\pi \hbar ^{2}}\left\{ h_{0}^{\left( 1\right) }\left( k\left\vert 
\mathbf{r}-\mathbf{r}^{\prime }\right\vert \right) -h_{0}^{\left( 1\right)
}\left( k\left\vert \mathbf{r}-\widetilde{\mathbf{r}}^{\prime }\right\vert
\right) \right\} ,  \label{G+0}
\end{equation}%
is the electron Green's function of Eq.~ (\ref{Schrod}) for the
semi-infinite half-space $\left( U_{0}\rightarrow \infty \right) $, $%
\widetilde{\mathbf{r}}^{\prime }=\left( \mathbf{\rho }^{\prime },-z^{\prime
}\right) ,$and $\varphi _{0}{\left( \mathbf{r}\right) }$ is given by Eq.~(%
\ref{phi0}). For small $g$ Eq.~(\ref{phi}) can be solved by perturbation
theory, i.e. in first approximation in ${g}$ the function ${\ \varphi \left( 
\mathbf{r}^{\prime }\right) }$ in the integral term should be replaced by ${%
\varphi }_{0}{\left( \mathbf{r}^{\prime }\right) .}$

\begin{figure}[tbp]
\includegraphics[width=8cm]{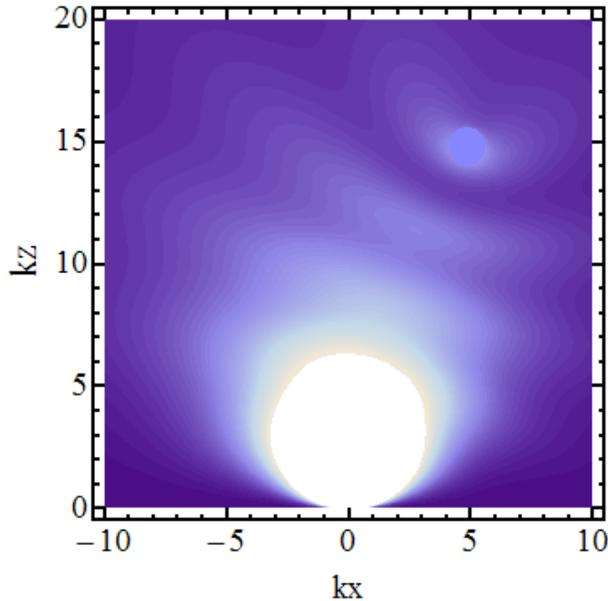}
\caption{Spatial distribution of the square modulus of the wave function in
the vicinity of the contact in the plane perpendicular to the interface
passing through the contact and the defect. Distances are given in units of
the inverse wave number \protect\cite{Avotina05}.}
\label{wf}
\end{figure}

For a short range potential ($kr_{D}$ $\ll 1$) the function ${\varphi \left( 
\mathbf{r}^{\prime }\right) }$ can be taken outside of integral in Eq.~(\ref%
{phi}) and the scattered wave function is written as \cite{Grenot} 
\begin{equation}
\varphi {\left( \mathbf{r}\right) }=\varphi _{0}{\left( \mathbf{r}\right) }%
+T\left( k\right) \varphi _{0}\left( \mathbf{r}_{0}\right) G_{0}^{+}\left( 
\mathbf{r,r}_{0};\varepsilon \right) ,  \label{phi_scat}
\end{equation}%
where 
\begin{equation}
T\left( k\right) =\frac{g}{1-g\int {d\mathbf{r}^{\prime }D}_{0}{\left(
\left\vert \mathbf{r}^{\prime }-\mathbf{r}_{0}\right\vert \right)
G_{0}^{+}\left( \mathbf{r}_{0}\mathbf{,r}^{\prime };\varepsilon \right) }}.
\label{T}
\end{equation}%
Note that Eq.(\ref{phi_scat}) is valid far from the defect ($\left\vert 
\mathbf{r}-\mathbf{r}_{0}\right\vert \gg r_{D}$) and the function $%
D_{0}\left( \left\vert \mathbf{r}^{\prime }-\mathbf{r}_{0}\right\vert
\right) $ must provide the convergence for the integral in the denominator
of Eq.~(\ref{T}) at $\mathbf{r}^{\prime }\rightarrow \mathbf{r}_{0}$. As is
well known, s-wave scattering is dominant for scattering by a short range
potential and the scattering matrix (\ref{T}) can be expressed by the s-wave
phase shift $\delta _{0}$ \cite{Avotina08mi} 
\begin{equation}
T\left( k\right) =\frac{i\pi \hbar ^{2}}{m^{\ast }k}\frac{e^{2i\delta _{0}}-1%
}{1+\frac{1}{2}\left( e^{2i\delta _{0}}-1\right) h_{0}^{\left( 1\right)
}\left( 2kz_{0}\right) }.  \label{T_d}
\end{equation}%
The effective $T$-matrix is an oscillatory function of the distance $z_{0}$
between the defect and the interface that results from repeated electron
scattering by the defect after its reflections from the interface. Figure~%
\ref{wf} illustrates the spatial distribution of the square modulus of the
wave function (\ref{phi_scat}) in the vicinity of the contact with the
defect placed at $\mathbf{r}_{0}=\left( 5,0,15\right) /k.$

\section{Friedel-like oscillations of the tunnel point contact conductance}

\subsection{Voltage dependence of the STM conductance}

In the case of a small transparency (\ref{t(k)}) the applied voltage drops
entirely over the barrier and the electrical potential can be chosen as a
step function $V(\mathbf{r})=V\,\Theta (-z).$ At zero temperature electrons
tunnel to the lower half-space when $eV>0$, and for $eV<0$ electrons can
tunnel only to available states in the upper half-space (Fig. \ref{Contact}).

The tunnelling current $I(V)=I^{\left( +\right) }(V)-I^{\left( -\right) }(V)$
is the difference between two currents flowing through the contact in
opposite directions. Each of them can be evaluated \ by means of the
probability current density $J_{k}^{\left( \pm \right) }(V)$ integrated over
a plane $z=\mathrm{const}$, and integrating over all directions of the
electron wave vector 
\begin{equation}
J_{k}^{(\pm )}(V)=\nu \left( \varepsilon \right) \int\limits_{-\infty
}^{\infty }d\mathbf{\rho }\Theta \left( \pm z\right) \left\langle \Theta
\left( \pm k_{z}\right) \func{Re}\varphi ^{\ast }(\mathbf{r})\widehat{v}%
_{z}\varphi (\mathbf{r})\right\rangle _{\varepsilon },  \label{Ik}
\end{equation}%
where $\nu \left( \varepsilon \right) $ is the electron density of states
for one spin direction, $\left\langle ...\right\rangle _{\varepsilon }$
denotes the average over an iso-energy surface $\varepsilon \left( \mathbf{k}%
\right) =\varepsilon $, 
\begin{equation}
\left\langle ...\right\rangle _{\varepsilon }=\left(
\int\limits_{\varepsilon \left( \mathbf{k}\right) =\varepsilon }\frac{dS_{%
\mathbf{k}}}{\left\vert \mathbf{v}\right\vert }\right)
^{-1}\int\limits_{\varepsilon \left( \mathbf{k}\right) =\varepsilon }\frac{%
dS_{\mathbf{k}}}{\left\vert \mathbf{v}\right\vert }...\;,  \label{average}
\end{equation}%
$dS_{\mathbf{k}}$ is an element of the iso-energy surface in $\mathbf{k}$%
-space, and $\widehat{\mathbf{v}}=\frac{1}{\hbar }\frac{\partial \varepsilon
\left( \widehat{\mathbf{k}}\right) }{\partial \widehat{\mathbf{k}}}$ is the
velocity operator. For a free-electron model of the energy spectrum $%
\widehat{v}_{z}=\frac{\hbar }{im^{\ast }}\frac{\partial }{\partial z}.$ A
voltage dependence of the current density $J_{k}^{\left( \pm \right) }(V)$ (%
\ref{Ik}) is defined by the dependence of the absolute value of the wave
vector for the incident on the contact electron $|\mathbf{k}\left( V\right)
|=\sqrt{k^{2}-2m^{\ast }\left\vert eV\right\vert /\hbar ^{2}}$.

The total current through the contact is%
\begin{gather}
{I(V)=}e\sum\limits_{\sigma =\pm 1}\int d\varepsilon {\left[ J_{k}^{\left(
+\right) }(V)f_{\mathrm{F}}\left( \varepsilon -eV\right) \left( {\newline
1-f_{\mathrm{F}}\left( \varepsilon \right) }\right) \right. -}  \notag \\
{\left. J_{k}^{\left( -\right) }(V)f_{\mathrm{F}}\left( \varepsilon \right)
\left( 1-f_{\mathrm{F}}\left( \varepsilon -eV\right) \right) \right] ,}
\label{I(V)}
\end{gather}%
where $f_{\mathrm{F}}\left( \varepsilon \right) $ is the Fermi function.

The current-voltage characteristics $I\left( V\right) $ is calculated by
substituting wave function (\ref{phi_scat}) into Eq.~(\ref{I(V)}) and taking
into account Eqs.~(\ref{phi0}) and (\ref{G+0}). Retaining only terms to
first order in $g$ (i.e. ignoring multiple scattering at the impurity site,
in Eq.~(\ref{phi_scat}) $T(k)\sim g$), and in the limit of low temperatures, 
$T=0$ , the conductance $G(V)=dI/dV$ can be written as \cite{Avotina05} 
\begin{gather}
G\left( \mathbf{r}_{0},V\right) =\frac{e^{2}\hbar ^{3}}{4\pi ^{3}\left(
m^{\ast }U_{0}\right) ^{2}}\iint \frac{d\mathbf{\rho }_{1}d\mathbf{\rho }_{2}%
}{f\left( \mathbf{\rho }_{1}\right) f\left( \mathbf{\rho }_{2}\right) }\times
\label{GV} \\
\left[ k_{\mathrm{F}}^{2}\widetilde{k}_{\mathrm{F}}^{4}F_{\widetilde{k}_{%
\mathrm{F}}}\left( \mathbf{\rho }_{1},\mathbf{\rho }_{2}\right)
-2\int\limits_{k_{\mathrm{F}}}^{\widetilde{k}_{\mathrm{F}}}k^{5}dkF_{k}%
\left( \mathbf{\rho }_{1},\mathbf{\rho }_{2}\right) \right] ,  \notag
\end{gather}%
where%
\begin{equation}
F_{k}\left( \mathbf{\rho }_{1},\mathbf{\rho }_{2}\right) =\left[ \frac{%
j_{1}\left( k\rho \right) }{k\rho }\right] ^{2}-\frac{4m^{\ast }gk}{\pi
\hbar ^{2}}\frac{j_{1}\left( k\rho \right) }{k\rho }\frac{z_{0}^{2}}{\lambda
_{1}\lambda _{2}}j_{1}\left( k\lambda _{1}\right) y_{1}\left( k\lambda
_{2}\right) ,
\end{equation}%
$\rho =\left\vert \mathbf{\rho }_{1}-\mathbf{\rho }_{2}\right\vert ,$ $%
\lambda _{1}=\sqrt{z_{0}^{2}+\left\vert \mathbf{\rho }_{0}-\mathbf{\rho }%
_{1}\right\vert ^{2}},$ $\lambda _{2}=\sqrt{z_{0}^{2}+\left\vert \mathbf{%
\rho }_{0}-\mathbf{\rho }_{2}\right\vert ^{2}},$ and $j_{l}(x)$ and $%
y_{l}(x) $ are the spherical Bessel functions, and 
\begin{equation}
\widetilde{k}_{\mathrm{F}}\left( V\right) =\sqrt{k_{\mathrm{F}}^{2}+2m^{\ast
}eV/\hbar ^{2}}  \label{k_F(V)}
\end{equation}
is the Fermi wave vector $k_{\mathrm{F}}$ accelerated by the potential
difference. In Eq.(\ref{GV}) for definiteness a positive sign of the bias is
chosen, $eV>0$.

If the contact radius $a\ll \lambda _{\mathrm{F}}$ ($\lambda _{\mathrm{F}%
}=1/k_{\mathrm{F}}$ is the Fermi wave length), the expression for the
conductance (\ref{GV}) can be simplified%
\begin{equation}
G\left( \mathbf{r}_{0},V\right) =G_{0}\left\{ q\left( \frac{eV}{\varepsilon
_{\mathrm{F}}}\right) -\widetilde{g}\frac{z_{0}^{2}}{r_{0}^{2}}\left[ \left( 
\frac{\widetilde{k}_{\mathrm{F}}}{k_{\mathrm{F}}}\right) ^{5}w\left( 
\widetilde{k}_{\mathrm{F}}r_{0}\right) -\left( \frac{\widetilde{k}_{\mathrm{F%
}}}{k_{\mathrm{F}}}\right) ^{7}v\left( \widetilde{k}_{\mathrm{F}%
}r_{0}\right) +v\left( k_{\mathrm{F}}r_{0}\right) \right] \right\} ,
\label{GVa->0}
\end{equation}%
where 
\begin{equation}
G_{0}=\left\vert t\left( k_{\mathrm{F}}\right) \right\vert ^{2}\frac{%
e^{2}\left( k_{\mathrm{F}}a\right) ^{4}}{36\pi \hbar }  \label{G0}
\end{equation}%
is the inherent conductance of the tunnel point contact, $r_{0}=\sqrt{%
z_{0}^{2}+\left\vert \mathbf{\rho }_{0}\right\vert ^{2}}$,%
\begin{equation}
q(x)=1+x-\frac{1}{3}x^{3},  \label{q}
\end{equation}%
\begin{equation}
w\left( x\right) =\frac{1}{x^{4}}\left[ \left( x^{2}-1\right) \sin 2x+2x\cos
2x\right] ,  \label{w}
\end{equation}%
\begin{equation}
v\left( x\right) =\frac{1}{x^{7}}\left[ 2x\left( 4x^{2}-7\right) \sin
2x+\left( 2x^{4}-14x^{2}+7\right) \cos 2x\right] ,
\end{equation}%
and 
\begin{equation}
\widetilde{g}=\frac{6m^{\ast }k_{\mathrm{F}}}{\pi \hbar ^{2}}g,  \label{g}
\end{equation}%
is the dimensionless constant of interaction.

Equation~(\ref{GVa->0}) describes the oscillations of the STM conductance as
a function of the distance $r_{0}$ between the STM tip and the subsurface
defect, and as a function of the bias $eV$. For distances between the
contact and the defect $r_{0}\gg \lambda _{\mathrm{F}}$ and $eV\ll
\varepsilon _{\mathrm{F}}$ the oscillatory dependence becomes sinusoidal 
\begin{equation}
G(\mathbf{r}_{0},V)-G_{0}\propto \frac{z_{0}^{2}}{r_{0}^{4}}\sin 2\widetilde{%
k}_{\mathrm{F}}r_{0}.\quad  \label{Gosc(V)}
\end{equation}%
Oscillations of the STM conductance as a function of the voltage, due to the
quantum interference caused by impurity scattering, were observed by Untiedt 
\textit{et al.} \cite{Untiedt}, and Ludoph \textit{et al.} \cite{Ludoph}.

\subsection{Determination of the defect positions}

Now we proceed to discuss whether this effect can be exploited
experimentally for three dimensional mapping of subsurface impurities. The
position of the defect in the plane parallel to the surface can be found
from an analysis of oscillatory pattern in the dependence $G\left( \mathbf{%
\rho }_{0}\right) .$ In the majority of cases the center of this pattern
corresponds to the tip position directly above the defect, $\mathbf{\rho }%
_{0}=0.$ A possible effect of the Fermi surface anisotropy is discussed in
the next section. Note that, in contrast to the case of surface defects, the
oscillations in the conductance (\ref{GVa->0}) are not periodic in the tip
distance $\mathbf{\rho }_{0}$ along the surface, but their period is defined
by the distance $r_{0}=\sqrt{\rho _{0}^{2}+z_{0}^{2}}$. Generally, the depth 
$z_{0}$ may be found by fitting the experimental data to the theoretical
dependence $G\left( \mathbf{\rho }_{0},z_{0}\right) $ (\ref{GVa->0}).
Figure~3 illustrates the oscillatory component of $G\left( \mathbf{r}%
_{0},V=0\right) $ as a function of $\rho _{0} $ for different choices of $%
z_{0}$. In this plot we have used the values for the constant $\widetilde{g}%
=1,$ the Fermi wave vector $k_{\mathrm{F}}=1.360$ \AA $^{-1}$ and the
interatomic distance $d=1.805$ \AA\ for Cu. Thus, the plots correspond to
defect positions in the third, fourth, and fifth layers below the Cu
surface. The dependencies $G\left( \rho \right) $ closely resemble the
observations by Quaas \textit{et al.} \cite{Quaas} for Co atoms embedded in
Cu(111).

For the determination of the defect depth $z_{0}$ one may use the
periodicity in phase $\vartheta =2\widetilde{k}_{\mathrm{F}}r_{0}$ of $G(%
\mathbf{r}_{0},V)$ (\ref{Gosc(V)}) at sufficiently large $r_{0}.$ According
to Eq.~(\ref{Gosc(V)}) at $V=0$ two sequential radii $\rho _{01}$ and $\rho
_{02}$, $\rho _{02}>\rho _{01}$, corresponding to neighboring maxima (or
minima) satisfy the obvious condition of periodicity, $\Delta \vartheta =2k_{%
\mathrm{F}}\left( \sqrt{\rho _{02}^{2}+z_{0}^{2}}-\sqrt{\rho
_{01}^{2}+z_{0}^{2}}\right) =2\pi .$ For known $k_{\mathrm{F}}$ it is a
simple algebraic equation for $z_{0},$ the solution of which is 
\begin{equation}
z_{0}=\frac{1}{2\pi k_{\mathrm{F}}}\sqrt{k_{\mathrm{F}}^{4}\left( \rho
_{02}^{2}-\rho _{01}^{2}\right) ^{2}-2\pi ^{2}k_{\mathrm{F}}^{2}\left( \rho
_{02}^{2}+\rho _{01}^{2}\right) +\pi ^{4}}.
\end{equation}%
Note that $k_{\mathrm{F}}\left( \rho _{02}-\rho _{01}\right) >\pi $ and the
radicand is positive. A second possibility of changing the product $%
\widetilde{k}_{\mathrm{F}}r_{0}$ is by varying the maximum value of the
electron wave vector by the applied voltage.

A first approach for determining the defect depth $z_{0}$ from the bias
dependence of the period of the Friedel-like oscillations of the STM
conductance was described by Kobayashi \cite{Kobayashi}. The depth $z_{0}$
can estimated by tracing the point $\rho _{0}$ while changing the bias
voltage $\left\vert eV\right\vert $, keeping the phase of the oscillations $%
\vartheta $ constant: $k_{\mathrm{F}}\sqrt{\rho _{0}^{2}+z_{0}^{2}}=%
\widetilde{k}_{\mathrm{F}}\left( V\right) \sqrt{\rho _{0}^{\prime
2}+z_{0}^{2}},$ where $\rho _{0}$ and $\rho _{0}^{\prime }$ are the
positions corresponding to two different bias voltages $V\rightarrow 0$ and $%
V_{2}=V$ but the same phase (for example, a fixed maximum) \cite{Kobayashi}.
The solution of mentioned above equation with $\widetilde{k}_{\mathrm{F}%
}\left( V\right) $ (\ref{k_F(V)}) gives $z_{0}$%
\begin{equation}
z_{0}=\sqrt{\frac{\varepsilon _{\mathrm{F}}\left( \rho _{0}^{2}-\rho
_{0}^{\prime 2}\right) -eV\rho _{0}^{\prime 2}}{eV}},
\end{equation}%
where $eV>0,$ $\rho _{0}>\rho _{0}^{\prime }.$

The method proposed in Ref.~\cite{Avotina05} has certain advantages. If the
STM tip is placed above the defect ($\left\vert \mathbf{\rho }%
_{0}\right\vert \ll z_{0}$) the conductance amplitude decreases with depth
of the defect as $z_{0}^{2}$, which gives hope to observe the defects at
sufficiently large distances below the surface. The depth of an impurity may
be derived from the $G(V)$ curve at $\mathbf{\rho }_{0}=0,$ which shows
oscillations in $eV$ with period $e\Delta V,$ and 
\begin{equation}
z_{0}=\frac{\pi }{k_{\mathrm{F}}-\widetilde{k}_{\mathrm{F}}\left( \Delta
V\right) }.
\end{equation}%
In a real experiment it is not necessary to observe a full period of $G(V)$
and, for example, a quarter of the period will be sufficient for the
determination of the defect depth \cite{Avotina05}. 
\begin{figure}[tbp]
\includegraphics[width=8cm]{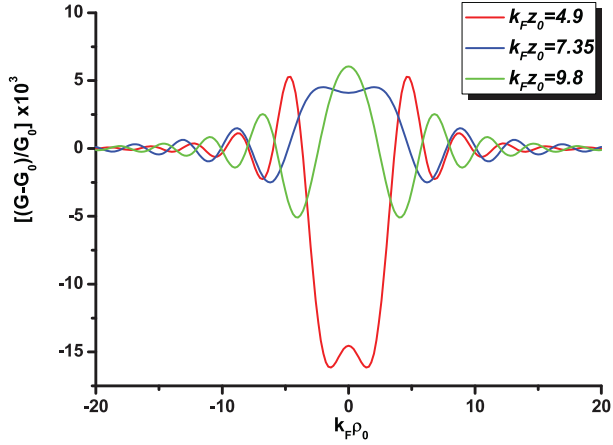}
\caption{Dependence of the normalized oscillatory part of the conductance on
the STM tip position for different depths $z_{0}$ of the defect below the
surface; $\mathbf{r}_{0}=\left( \protect\rho _{0},0,z_{o}\right) ,\widetilde{%
g}=1$.}
\end{figure}

\section{Signature of the Fermi surface anisotropy}

In most metals the dispersion relation for the charge carriers is a
complicated anisotropic function of momentum. This leads to anisotropy of
the various kinetic characteristics \cite{LAK}. Particularly, as shown in
Ref.~\cite{Kosevich}, the current spreading may be strongly anisotropic in
the vicinity of a point-contact. This effect influences the way the
point-contact conductance depends on the position of the defect. For
example, in the case of a Au(111) surface the `necks' in the Fermi surface
(FS) should cause a defect to be invisible when probed exactly from above.

Qualitatively, the wave function of electrons injected by a point contact
for arbitrary FS $\varepsilon \left( \mathbf{k}\right) =\varepsilon _{%
\mathrm{F}}$ has been analyzed by A. Kosevich \cite{Kosevich}. He noted that
at large distances from the contact the electron wave function for a certain
direction $\mathbf{r}$ is defined by those points on the FS for which the
electron group velocity is parallel to $\mathbf{r.}$ Unless the entire FS is
convex there are several such points. The amplitude of the wave function
depends on the Gaussian curvature $K$ in these points, which can be convex $%
\left( K>0\right) $ or concave $\left( K<0\right) $. The parts of the FS
having different signs of curvature are separated by lines of $K=0$
(inflection lines). In general there is a continuous set of electron wave
vectors for which $K=0$. The electron flux in the directions having zero
Gaussian curvature exceeds the flux in other directions \cite{Kosevich}.

Electron scattering by defects in metals with an arbitrary FS can be
strongly anisotropic \cite{LAK}. Generally, the wave function of the
electrons scattered by the defect consists of several superimposed waves,
which travel with different velocities. In the case of an open FS there are
directions along which the electrons cannot move at all. Scattering events
along those directions occur only if the electron is transferred to a
different sheet of the FS \cite{LAK}.

In this section we analyze the effect of anisotropy of the FS to the signals
for determination of the position of a defect below a metal surface by use
of a STM. We show below that the amplitude and the period of the conductance
oscillations are defined by the local geometry of the FS, namely by those
points for which the electron group velocity is directed along the radius
vector from the contact to the defect. General results are illustrated for
the FS of noble metals.

At first we do not specify the specific form of the dependence $\varepsilon
\left( \mathbf{k}\right) $, except that it satisfies the general condition
of point symmetry $\varepsilon \left( \mathbf{k}\right) =\varepsilon \left( -%
\mathbf{k}\right) $. In the reduced zone scheme a given vector $\mathbf{k}$
identifies a single point within the first Brillouin zone. As for isotropic
FS, the electron wave function $\psi \left( \mathbf{r}\right) $ in the metal
with an arbitrary dispersion relation can be found at $U_{0}\rightarrow
\infty $ by using the method described in Sec.III. The boundary conditions
for the transmitted wave function $\varphi _{0}\left( \mathbf{r}\right) $
have the same form as Eqs. (\ref{equal2}), (\ref{jump2}) in which the
function $t\left( \mathbf{k}\right) $ must be replaced by%
\begin{equation}
t\left( \mathbf{k}\right) =\frac{1}{U_{0}}\left[ \int dz\varepsilon \left( 
\mathbf{\kappa ,}\frac{\partial }{i\partial z}\right) \left(
e^{ik_{z}\left\vert z\right\vert }-e^{-ik_{z}^{\mathrm{ref}}\left\vert
z\right\vert }\right) \right] _{z=-0}.  \label{t_arb}
\end{equation}%
For the model of free electrons \ Eq.(\ref{t_arb}) transforms into Eq.(\ref%
{t(k)}). The components of vector $\mathbf{k}$ perpendicular to the
interface for electrons incident on the contact, $k_{z}\left( \mathbf{\kappa 
},\varepsilon \right) $, and reflected from the contact, $k_{z}^{\mathrm{ref}%
}\left( \mathbf{\kappa },\varepsilon \right) $, are related by conditions of
conservation of the energy $\varepsilon $ and the tangential component $%
\mathbf{\kappa }$ of the wave vector 
\begin{equation}
\varepsilon \left( \mathbf{\kappa }^{\mathrm{in}},k_{z}\right) =\varepsilon
\left( \mathbf{\kappa }^{\mathrm{ref}},k_{z}^{\mathrm{ref}}\right)
=\varepsilon ;\quad \mathbf{\kappa }^{\mathrm{in}}=\mathbf{\kappa }^{\mathrm{%
ref}}\equiv \mathbf{\kappa }.  \label{pz_ref}
\end{equation}%
The wave function scattered by the defect is defined by the general relation
(\ref{phi}).

General expressions for the STM conductance into a metal having an arbitrary
FS one can find in Ref.~\cite{Avotina06}. Here, we present simplified
asymptotic expressions for the oscillatory part of the conductance $\Delta
G_{osc}^{\mathrm{arb}}\left( \mathbf{r}_{0}\right) $ (the difference between
the total conductance and its value in the absence of the defect) which are
valid for large distances between the contact and the defect , $r_{0}\gg
\lambda _{\mathrm{F}}$ \cite{Avotina06}, 
\begin{gather}
\Delta G_{\mathrm{osc}}^{\mathrm{arb}}\left( \mathbf{r}_{0}\right) =\frac{%
2ge^{2}a^{4}z_{0}^{2}}{\hbar r_{0}^{4}}\nu \left( \varepsilon _{\mathrm{F}%
}\right) \left\langle \left\vert t\left( \mathbf{k}\right) \right\vert
^{2}\Theta \left( v_{z}\right) \right\rangle _{\varepsilon _{\mathrm{F}%
}}\cdot  \label{G_arbit} \\
\sum_{s,s^{\prime }}\frac{1}{\sqrt{\left\vert K\left( \mathbf{k}_{0s}\right)
K\left( \mathbf{k}_{0s^{\prime }}\right) \right\vert }}\sin (h\left( \mathbf{%
k}_{0s}\right) r_{0}+\phi _{s})\cos \left(h\left( \mathbf{k}_{0s^{\prime
}}\right) r_{0}+\phi _{s^{\prime }}\right) .  \notag
\end{gather}%
All functions of the wave vector in Eq.~(\ref{G_arbit}) are taken at the
points of the FS for which the electron group velocity $\mathbf{v}_{0}$ is
parallel to the vector $\mathbf{r}_{0}=r_{0}\mathbf{n}_{0}$, $h\left(
\varepsilon _{\mathrm{F}},\mathbf{k}_{0}\right) =\mathbf{k}_{0}\mathbf{n}%
_{0},$ $\mathbf{k}_{0}$ is the wave vector corresponding to the point on the
FS, in which $\mathbf{v}_{0}\mathbf{\mathbf{\parallel }n}_{0}.$ The function 
$h\left( \varepsilon _{\mathrm{F}},\mathbf{k}_{0}\right) $ is well known in
the differential geometry as the support function of the surface $%
\varepsilon \left( \mathbf{k}\right) =\varepsilon _{\mathrm{F}}$ \cite%
{Aminov}. If the curvature of the FS changes sign, there is more than one
point $\mathbf{k}_{0s}$ ($s=1,2...$) for which $\mathbf{v}_{0s}\parallel 
\mathbf{n}_{0}$. It may also occur that for given directions of the vector $%
\mathbf{r}_{0}$ $\mathbf{v}\nparallel \mathbf{n}_{0}$ for all points on the
FS, and the electrons cannot propagate along these directions \cite{LAK}.
For such $\mathbf{r}_{0}$ the oscillatory part of the conductance is zero.

In Eq.~(\ref{G_arbit}) 
\begin{equation}
\phi =\frac{\pi }{4}\text{sgn}\left( \frac{\partial ^{2}k_{z}^{\left(
+\right) }}{\partial k_{x}^{2}}\right) \left( 1+\text{sgn}K\left( \mathbf{k}%
_{0}\right) \right) ,  \label{lam}
\end{equation}%
$\left\langle \dots \right\rangle _{\varepsilon _{\mathrm{F}}}$ \ is defined
by Eq.~(\ref{average}), $k_{z}^{\left( +\right) }=k_{z}^{\left( +\right)
}\left( k_{x},k_{y},\varepsilon _{\mathrm{F}}\right) $ in the point defined
by the direction of the vector $\mathbf{n}_{0}$ in $\mathbf{k}$-space, $%
K\left( \mathbf{k}_{0}\right) \neq 0$ is the Gaussian curvature of the FS, 
\begin{equation}
K\left( \varepsilon _{\mathrm{F}},\mathbf{k}_{0}\right) =\frac{\hbar ^{2}}{%
\left\vert \mathbf{v}_{0}\right\vert ^{2}}\sum_{i,j=x,y,z}A_{ik}n_{0i}n_{0j%
\mathbf{\ }},  \label{Gaus0}
\end{equation}%
$A_{ij}=\frac{\partial \det \left( \mathbf{m}^{-1}\right) }{\partial
m_{ij}^{-1}\left( \mathbf{k}\right) }$ is the algebraic adjunct of the
element $m_{ij}^{-1}\left( \mathbf{k}\right) =\frac{1}{\hbar ^{2}}\frac{%
\partial ^{2}\varepsilon }{\partial k_{i}\partial k_{j}}$ of the inverse
mass matrix $\mathbf{m}^{-1}$.

The Eq.(\ref{G_arbit}) is valid, if curvature $K\neq 0.$ For those points at
which $K$ the amplitude of the electron wave function in a direction of zero
Gaussian curvature is larger than for other directions. This results in an
enhanced current flow near the cone surface defined by the condition $K=0$ 
\cite{Kosevich,Avotina06}. If the FS is open, there are directions along
which the electron flow is absent. These properties of the wave function
manifest itself in an oscillatory part of the conductance (\ref{G_arbit}):
1) The amplitude of oscillations is maximal if the direction from the
contact to the defect corresponds to the electron velocity belonging to an
inflection line. 2) There are no conductance oscillations, $\Delta G_{osc}^{%
\mathrm{arb}}=0,$ if this direction belong to cones, in which the electron
motion is forbidden.

For an ellipsoidal FS the Schr\"{o}dinger equation can, in fact, be solved
exactly in the limit $a\rightarrow 0,$ $U_{0}\rightarrow \infty $ and the
conductance of the contact can be found for arbitrary distances between the
contact and the defect. For this FS the dependence of the electron energy $%
\varepsilon $ on the wave vector $\mathbf{k}$ is given by relation, 
\begin{equation}
\varepsilon \left( \mathbf{k}\right) =\frac{\hbar ^{2}}{2}\sum_{i,j=x,y,z}%
\frac{k_{j}k_{i}}{m_{ij}};  \label{energy}
\end{equation}%
were $k_{i}$ are the components of the electron wave vector $\mathbf{k}$, $%
1/m_{ij}$ are constants representing the components of the inverse effective
mass tensor $\mathbf{m}^{-1}$.

Accurate to within first order in $g$ (i.e. ignoring multiple scattering at
the impurity site), the conductance in the limit $V\rightarrow 0$ \cite%
{Avotina06} is given by 
\begin{equation}
G^{\mathrm{ell}}\left( \mathbf{r}_{0}\right) =G_{0}^{\mathrm{ell}}\left[ 1-%
\frac{6g\left( 2\varepsilon _{\mathrm{F}}\right) ^{3/2}}{\pi \hbar ^{5}\sqrt{%
m_{zz}}\det \left[ \mathbf{m}^{-1}\right] }\left( \frac{z_{0}}{h\left( 
\mathbf{k}_{0}\right) r_{0}}\right) ^{2}w\left( h\left( \mathbf{k}%
_{0}\right) r_{0}\right) \right]  \label{G_ell}
\end{equation}%
where $G_{0}^{\mathrm{ell}}$ in Eq.~(\ref{G_ell}) is the conductance in the
absence of a defect $\left( g=0\right) $ \cite{Avotina06}: 
\begin{equation}
G_{0}^{\mathrm{ell}}=\frac{2e^{2}a^{4}\varepsilon _{\mathrm{F}}^{3}}{9\pi
\hbar ^{3}U_{0}^{2}\sqrt{m_{zz}\det \left[ \mathbf{m}^{-1}\right] }}.
\label{G_c}
\end{equation}%
\begin{equation}
h\left( \mathbf{k}_{0}\right) =\frac{1}{\hbar }\left( \frac{2\varepsilon _{%
\mathrm{F}}}{\det \left[ \mathbf{m}^{-1}\right] }%
\sum_{i,j=x,y,z}A_{ij}n_{0i}n_{0j}\right) ^{1/2},
\end{equation}%
$w\left( kr\right) $ is given by Eq.~(\ref{w}).

\begin{figure}[tbp]
\includegraphics[width=8cm,angle=0]{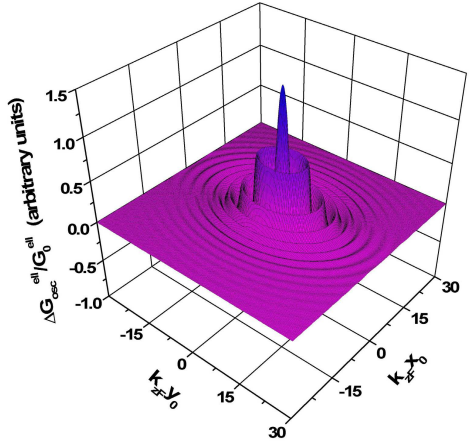}
\caption{Dependence of the oscillatory part of the conductance, $\Delta
G_{osc}^{\mathrm{ell}}$, as a function of the position of the defect $%
\mathbf{\protect\rho }_{0}$ in the plane $z=z_{0}$. The shape of the FS (%
\protect\ref{energy}) is defined by the mass ratios $m_{x}/m_{z}$=1, $%
m_{y}/m_{z}=3$, and the long axis of the ellipsoid is rotated by $\protect%
\pi /4$ around the $x-$axis, away from the $y-$axis. The coordinates are
measured in units $1/k_{z\mathrm{F}}$ (\protect\ref{k_zF}) and the defect
sits at $z_{0}=5$ \protect\cite{Avotina06}. }
\label{fig-ell-deltaG}
\end{figure}

The center of the oscillation pattern in the conductance $G^{\mathrm{ell}%
}\left( \mathbf{r}_{0}\right) $ as the function of the tip position $\mathbf{%
\rho }_{0}$ corresponds to $\mathbf{\rho }_{0}=\mathbf{\rho }_{00}$ with
respect to the point contact at $\mathbf{r}=0$, where 
\begin{equation}
\mathbf{\rho }_{00}=z_{0}\left( \frac{m_{zz}}{m_{zx}},\frac{m_{zz}}{m_{zy}}%
\right) .  \label{x0y0}
\end{equation}%
The support function $h$ for such tip position, 
\begin{equation}
\mathbf{k}_{0}\mathbf{n}_{00}\equiv k_{z\mathrm{F}}=\frac{1}{\hbar }\sqrt{%
2\varepsilon _{\mathrm{F}}m_{zz}};  \label{k_zF}
\end{equation}%
corresponds to the extremal value of the chord $2k_{z\mathrm{F}}$ of the FS
in the direction normal to the interface, $\mathbf{n}_{00}$ is the unit
vector in the direction of the vector $\mathbf{r}_{00}=\left( \mathbf{\rho }%
_{00},z_{0}\right) $. Figure~\ref{fig-ell-deltaG} shows that $\Delta
G_{osc}^{\mathrm{ell}}=G^{\mathrm{ell}}-G_{0}^{\mathrm{ell}}$ is an
oscillatory function of the defect position $\mathbf{\rho }_{0}$ that
reflects the ellipsoidal form of the FS and the oscillations are largest
when the contact is placed in the position $\mathbf{\rho }_{00}$, defined by
Eq.~(\ref{x0y0}).

In deriving Eq.~(\ref{G_arbit}) it has been assumed that $eV\rightarrow 0.$
For finite voltage, but $eV\ll \varepsilon _{\mathrm{F}}$, all functions of
the energy $\varepsilon $ in Eq.~(\ref{G_arbit}) can be taken at $%
\varepsilon =\varepsilon _{\mathrm{F}}$, except $h\left( \varepsilon ,%
\mathbf{k}_{0}\right) $ in the oscillatory functions. When $eV\ll
\varepsilon _{\mathrm{F}}$, 
\begin{equation}
h\left( \varepsilon _{\mathrm{F}}+eV,\mathbf{k}_{0}\right) \approx h\left(
\varepsilon _{\mathrm{F}},\mathbf{k}_{0}\right) +\frac{\partial h}{\partial
\varepsilon _{\mathrm{F}}}eV,\quad \frac{\partial h}{\partial \varepsilon _{%
\mathrm{F}}}\sim k_{\mathrm{F}}\frac{eV}{\varepsilon _{\mathrm{F}}}
\end{equation}%
and when the product $(eV/\varepsilon _{\mathrm{F}})k_{\mathrm{F}}r_{0}\gg 1$
clearly the conductance (\ref{G_arbit}) is an oscillatory function of the
voltage $V$. The periods of the oscillations are defined by the energy
dependence of the function $h\left( \varepsilon ,\mathbf{k}_{0}\right) $.
The results obtained properly describe the total conductance at $eV\ll
\varepsilon _{\mathrm{F}}$ and also can be used for the analysis of the
periods of the oscillations at $eV\leq \varepsilon _{\mathrm{F}}.$

Further calculations require information about the actual shape of the FS, $%
\varepsilon \left( \mathbf{k}\right) =\varepsilon _{\mathrm{F}}.$ In Ref.~%
\cite{Avotina06} a model FS in the form of a corrugated cylinder was
considered. Using this model, for which analytical dependencies of the
conductance on defect position can be found, the manifestation of common
features of FS geometries in the conductance oscillations was described: the
anisotropy of convex parts (`bellies'), the changing in sign of the
curvature (inflection lines), and the presence of open directions (`necks').

In Ref.~\cite{Avotina08nm} a numerical analysis of the conductance
oscillation pattern was made for the noble metals copper, silver and gold on
the basis of Eq.~(\ref{G_arbit}). The parameterization of the FS was taken
from \cite{FS}, 
\begin{eqnarray}
\varepsilon (\mathbf{k}) &=&\alpha \left[ -3+\text{cos}\frac{k_{x}a}{2}\text{%
cos}\frac{k_{y}a}{2}+\text{cos}\frac{k_{y}a}{2}\text{cos}\frac{k_{z}a}{2}+%
\text{cos}\frac{k_{z}a}{2}\text{cos}\frac{k_{x}a}{2}+\right.  \label{E(k)} \\
&&\left. r\left( -3+\text{cos}k_{x}a+\text{cos}k_{y}a+\text{cos}%
k_{z}a\right) \right],  \notag
\end{eqnarray}%
which is accurate up to 99\%. The values for the constants are $r=0.0995$, $%
\varepsilon /\alpha $ $=3.63$, and $a$ is different for each metal. For
copper, silver, and gold $a=0.361$nm, $a=0.408$nm, and $a=0.407$nm,
respectively. The Fermi energy of copper is 7.00eV, for silver 5.49eV and
for gold it is 5.53eV.

\begin{figure}[tbp]
\includegraphics[width=8cm]{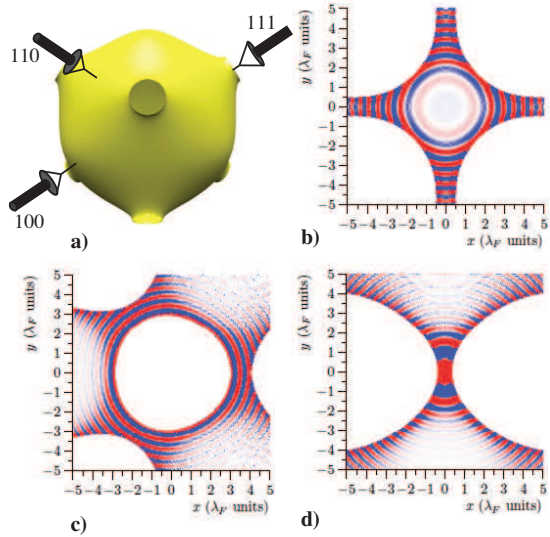}
\caption{\textbf{a)} Fermi surface described by Eq.~(\protect\ref{E(k)})
relative to the contact axis for three principal lattice orientations. 
\textbf{b)} Gray scale plot of the tunnelling point contact conductance $G$
as a function of the contact position for a defect at the origin, at a depth
of $5\protect\lambda _{F}$ and for a (100) surface plane; the $x$ and $y$
directions each correspond to $\left\langle 100\right\rangle $ directions. 
\textbf{c)} Same plot for a (111) surface orientation; the $x$ and $y$
directions correspond to $\left[ 11\bar{2}\right] $and $\left[ 1\bar{ 1}0%
\right] $ directions, respectively. \textbf{d)} Same plot for a (110)
surface orientation; the $x$ and $y$ directions correspond to $\left[ 001%
\right] $and $\left[ 1\bar{1} 0\right] $ directions, respectively 
\protect\cite{Avotina08nm}.}
\label{fig_nm}
\end{figure}

The results of computations for three crystallographic orientations are
presented in Fig.~\ref{fig_nm}. All distances in Fig.~\ref{fig_nm} are given
in units of $\lambda _{\mathrm{F}},$ which for copper is 0.46nm, and for
silver and gold it is 0.52nm. For each of the surface orientations the
graphs have the symmetries of that particular orientation of the FS. In all
figures 'dead' regions can be seen, for which the conductance of the contact
is equal to its value without the defect, showing no conductance
oscillations. These regions originate from the 'necks' of the FS and their
edges are defined by the inflection lines. For all orientations of the metal
surface the defect position in the plane of the surface corresponds to a
center of symmetry. The appearance of 'dead' regions depends on the depth of
the defect, which can be estimated in the following way: The orientations of
the 'neck' axes define the axes of the cones with an opening angle $2\gamma $%
, in which there are no scattered electrons. Vertexes of the cones coincide
with the defect. The radius $R$ of the central 'dead' region, $R=z_{0}\tan
\left( \gamma \right) ,$ is proportional to the depth of the defect \cite%
{Quaas}.

The possibility of visualizing the Fermi surface of Cu in real space by
investigation of interference patterns caused by subsurface Co atoms has
been demonstrated by Weismann \textit{et al.} \cite{Weismann}.

\section{Subsurface magnetic defects}

\subsection{Kondo impurity}

In the case of a magnetic defect at low temperatures ($T\ll T_{K}$, where $%
T_{K}$ is the Kondo temperature) the Kondo resonance results in a dramatic
enhancement of the effective electron-impurity interaction \cite{Abrikosov}
and perturbation methods become inapplicable. Kondo correlations give rise
to a sharp resonance in the density of states at the energy $\varepsilon (%
\mathbf{k})=\varepsilon _{\mathrm{K}}$ near the Fermi level. For $%
\varepsilon (\mathbf{k})\rightarrow \varepsilon _{\mathrm{K}}$ the effective
electron scattering cross section acquires a maximum value corresponding to
the Kondo phase shift $\delta _{0\mathrm{K}}=\pi /2$ \cite{Abrikosov}. In
this case, multiple scattering needs to be taken into account, even for a
single defect, because of electron reflection by the metal surface.

In this subsection the conductance is expressed by means of a s-wave
scattering phase shift $\delta _{0}$. The results describe the influence to
the conductance of multiple scattering of the electrons, which results in
the appearance of harmonics in the dependencies of $G$ on the applied
voltage, and on the distance between the contact and the defect. The
analysis of the non-monotonic voltage dependence of the conductance is
applied specifically to the interesting problem of Kondo scattering, using
an appropriate phase shift \cite{Shift}: 
\begin{equation}
\delta _{0}(k)=\left[ \frac{\pi }{2}-\tan ^{-1}\left( \frac{\varepsilon ({%
\mathbf{k}})-\varepsilon _{\mathrm{K}}}{T_{\mathrm{K}}}\right) \right]
+\delta _{0\mathrm{D}}.  \label{d0}
\end{equation}%
The first term in Eq.~(\ref{d0}) describes the resonant scattering on a
Kondo impurity level $\varepsilon _{\mathrm{K}}$ ($T_{\mathrm{K}}$ is the
Kondo temperature). For a non-magnetic impurity this term is absent. The
second term $\delta _{0\mathrm{D}}$ takes into account the usual potential
scattering.

Taking Eq.~(\ref{phi_scat}) for the wave function of a spherical Fermi
surface and Eq.~(\ref{T_d}) for the scattering matrix makes it possible to
find the differential conductance, $G=dI/dV$, of the tunnel point contact in
the approximation of s-wave scattering. For $\left\vert eV\right\vert
<\varepsilon _{\mathrm{F}}$ and for $eV>0,$ $G(V)$ is given by \cite%
{Avotina08mi} 
\begin{equation}
G(V)=G_{0}\left[ q(V)+\left( \frac{\widetilde{k}_{\mathrm{F}}}{k_{\mathrm{F}}%
}\right) ^{4}\Phi (\widetilde{k}_{\mathrm{F}})-\frac{2}{k_{\mathrm{F}}^{6}}%
\int\limits_{k_{\mathrm{F}}}^{\widetilde{k}_{\mathrm{F}}}dkk^{5}\Phi \left(
k\right) \right] ,  \label{G>}
\end{equation}%
and for $eV<0$, 
\begin{equation}
G(V)=G_{0}\left[ q(V)+\left( \frac{\widetilde{k}_{\mathrm{F}}}{k_{\mathrm{F}}%
}\right) ^{2}\Phi (\widetilde{k}_{\mathrm{F}})-\frac{4}{k_{\mathrm{F}}^{6}}%
\int\limits_{k_{\mathrm{F}}}^{\widetilde{k}_{\mathrm{F}}}dkk^{3}\left( k^{2}-%
\frac{2m^{\ast }eV}{\hbar ^{2}}\right) \Phi \left( k\right) \right] .
\label{G<}
\end{equation}%
Here $G_{0}$ is given by Eq.~(\ref{G0}), $\widetilde{k}_{\mathrm{F}}=\sqrt{%
k_{\mathrm{F}}^{2}+2m^{\ast }eV/\hbar ^{2}},$ and 
\begin{gather}
{\Phi \left( k\right) =F^{-1}\sin \delta _{0}\frac{z_{0}^{2}}{r_{0}^{2}}%
\left[ 12j_{1}\left( kr_{0}\right) \right. \left( -y_{1}\left( kr_{0}\right)
\cos \delta _{0}\right. +}  \label{Fi(k)} \\
{\newline
\left. \left\{ j_{1}\left( kr_{0}\right) \left( j_{0}\left( 2kz_{0}\right)
-1\right) +y_{0}\left( 2kz_{0}\right) y_{1}\left( kr_{0}\right) \right\}
\sin \delta _{0}\right) +}  \notag \\
{\newline
\left. 6\left( 1-j_{0}\left( 2kz_{0}\right) \right) \left( kr_{0}\right)
^{-4}\left( 1+\left( kr_{0}\right) ^{2}\right) \sin \delta _{0}\right] ,} 
\notag
\end{gather}%
\begin{eqnarray}
F &=&1+2\sin \delta _{0}\times  \label{D} \\
&&\left[ \left( \frac{1}{2\left( 2kz_{0}\right) ^{2}}-j_{0}(2kz_{0})\right)
\sin \delta _{0}-y_{0}(2kz_{0})\cos \delta _{0}\right] ,  \notag
\end{eqnarray}%
$\delta _{0}\left( k\right) $ is s-wave phase shift \ref{d0}, and $j_{l}(x)$
and $y_{l}(x)$ are the spherical Bessel functions.

At low voltage the conductance can be expressed by an expansion in the small
parameter $1/\left( k_{\mathrm{F}}z_{0}\right) <1$, 
\begin{gather}
{G(0)=G_{0}\left\{ 1+12\frac{z_{0}^{2}}{r_{0}^{2}}\frac{1}{\left( k_{\mathrm{%
F}}r_{0}\right) ^{2}}\sum\limits_{n=1}^{\infty }\left( -1\right) ^{n}\frac{%
\sin ^{n}\delta _{0}}{\left( 2k_{\mathrm{F}}z_{0}\right) ^{n-1}}\times
\right. }  \label{G(0)} \\
{\newline
\left[ \frac{1}{2}\left( 1-\frac{1}{\left( k_{\mathrm{F}}r_{0}\right) ^{2}}%
\right) \sin \left( 2k_{\mathrm{F}}\left( r_{0}+\left( n-1\right)
z_{0}\right) +n\delta _{0}\right) +\right. }  \notag \\
{\newline
\left. \left. \frac{1}{k_{\mathrm{F}}r_{0}}\cos \left( 2k_{\mathrm{F}}\left(
r_{0}+\left( n-1\right) z_{0}\right) +n\delta _{0}\right) \right] \right\} }
\notag
\end{gather}%
The second term in Eq.~(\ref{G(0)}) gives the sum over $n$ scattering events
by the defect and $n-1$ reflections by the surface. If we keep only the term
for $n=1$ Eq.~(\ref{G(0)}) reduces to the result obtained by perturbation
theory in Sec.~3 above, which is valid for $\delta _{0}\simeq -gm^{\ast }k_{%
\mathrm{F}}/2\pi \hbar ^{2}\ll 1.$ The arguments of the sine and cosine
functions in Eq.~(\ref{G(0)}) correspond to the phase that the electron
accumulates while moving along semiclassical trajectories.

The voltage dependence of the conductance is not symmetric around $V=0$.
This asymmetry arises from the dependencies of the phase shift $\delta _{0}(%
\widetilde{k}_{\mathrm{F}})$ (\ref{d0}) and the absolute value of the wave
vector $\widetilde{k}_{\mathrm{\ F}}=\sqrt{k_{\mathrm{F}}^{2}+2m^{\ast
}eV/\hbar ^{2}}$ on the sign of $eV$. The physical origin of this asymmetry
comes from the fact that the scattering amplitude depends on the electron
energy in the lower half-space (see Fig.~\ref{Contact}), where the defect is
situated. This energy is different for different directions of the current.

It is interesting to observe that the sign of the Kondo anomaly depends on
the distance between the contact and the defect $r_{0}$. This distance in
combination with the value of the wave vector $\widetilde{k}_{\mathrm{F}}$
determines the period of oscillation of $G(V)$. If the bias $eV_{\mathrm{K}}$
coincides with a maximum in the oscillatory part of conductance the sign of
the Kondo anomaly is positive and vice versa, a negative sign of the Kondo
anomaly is found at a minimum in the periodic variation of $G(V).$

\begin{figure}[tbp]
\includegraphics[width=8cm]{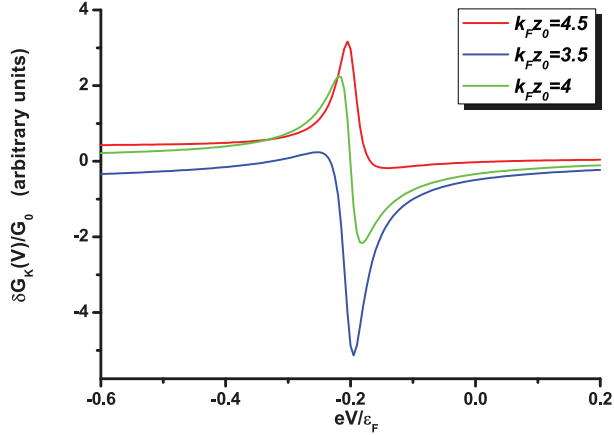}
\caption{Difference $\protect\delta G_{\mathrm{K}}(V)/G_{0}$ between the
voltage dependencies of the conductance for a magnetic and a non-magnetic
impurity. The parameters $\protect\varepsilon _{\mathrm{K}}=0.9\protect%
\varepsilon _{\mathrm{F}},$ $T_{\mathrm{K}}=0.01\protect\varepsilon _{%
\mathrm{F}},$ and $\protect\delta _{0\mathrm{D}}=0.1$ in Eq.\protect\ref{d0}
are used \protect\cite{Avotina08mi}.}
\label{Kondo}
\end{figure}

The Fig.\ref{Kondo}~ shows the difference $\delta G_{\mathrm{K}%
}(V)/G_{0}=(G_{m}-G_{n})/G_{0}$ between voltage dependencies for a magnetic $%
G_{m}$ and a non-magnetic $G_{n}$ impurity, having the same potential
scattering strength. The plots in Fig.\ref{Kondo}~ demonstrate the evolution
of the shape of the Kondo anomaly for several values of the distance between
the contact and the impurity, placed on the contact axis. The change of
distance changes the periodicity of the normal-scattering oscillations which
leads to a changing of sign in the Kondo signal. A similar dependence of the
differential conductance with the distance between an STM tip and an adatom
on the surface of a metal has been obtained theoretically in Refs.~\cite%
{Zawad,Lin} in terms of a Anderson impurity Hamiltonian \cite{Anderson}.
Note that we obtain a Fano-like shape of the Kondo resonance in the
framework of a single-electron approximation \cite{Avotina08mi}, while in
Refs.~\cite{Zawad,Lin} many-body effects were taken into account.

\subsection{Magnetic cluster}

In this subsection we consider the influence on the conductance of a tunnel
point-contact between magnetic and non-magnetic metals of a defect having an
unscreened magnetic moment, in a spin-polarized scanning tunnelling
microscope (SP-STM) geometry \cite{Bode}. A magnetic cluster is assumed to
be embedded in a non-magnetic metal in the vicinity of the contact. As first
predicted by Frenkel and Dorfman \cite{Frenkel} particles of a ferromagnetic
material are expected to organize into a single magnetic domain below a
critical particle size (a typical value for this critical size for Co is
about 35nm). Depending on the size and the material, the magnetic moments of
such particles can be $\mu _{\mathrm{eff}}\sim 10^{2}-10^{5}\mu _{\mathrm{B}%
} $ \cite{Vonsovsky}.

Generally, the moment $\mathbf{\mu }_{\mathrm{eff}}$ of the cluster in a
non-magnetic metal is free to choose an arbitrary direction. This direction
can be held fixed by an external magnetic field $\mathbf{H}$, the value of
which is estimated as $H\simeq T/\mu _{\mathrm{eff}}$, where $T$ is the
temperature (see, for example, Ref.~\cite{Vonsovsky}). For $\mu _{\mathrm{eff%
}}\simeq 10^{2}\mu _{\mathrm{B}}$ and $T\sim 1$K the field $H$ is of the
order of $0.01$T. If $H$ is much larger than the magnetocrystalline
anisotropy field of the magnetic STM tip, the direction of the external
magnetic field controls the direction of the cluster magnetic moment but its
influence on the spin-polarization of the tunnel current is negligible. In
this case the magnetic moment $\mathbf{\mu }_{\mathrm{eff}}$ of the cluster
is 'frozen' by the field $H$ and the problem becomes a stationary one.

If the external magnetic field is sufficiently weak and the radius of the
electron trajectories $r_{\mathrm{H}}=\hbar ck_{\mathrm{F}}/eH$ is much
larger than the distance between the contact and the cluster $r_{0}$, the
effects of modulation of the tunnel current due to electron spin precession 
\cite{Jedema} and trajectory magnetic effects \cite{Avotina07} are
negligible.

The geometry of a SP-STM experiment can be described in the framework of the
model presented in Fig.~\ref{Contact}, in which the half-space $z<0$ is
taken up by a ferromagnetic conductor with magnetization $\mathbf{M}$. In
Ref.~\cite{Avotina09} the direction of the vector $\mathbf{M}$ , which
defines the direction of the polarization of tunnel current is chosen along
the $z$-axis. In real SP-STM the polarization of STM current is defined by
the magnetization of the last atom of the tip \cite{Bode}. A magnetization
oriented along the contact axis can be obtained, for example, for a
Fe/Gd-coated W STM tip \cite{Kub}.

The interaction potential $\widehat{D}(\mathbf{r})$ of the electrons with
the cluster is a matrix consisting of two parts 
\begin{equation}
\widehat{D}(\mathbf{r})=\left( g\widehat{I}+\frac{1}{2\mu _{\mathrm{B}}}J%
\mathbf{\ \mu }_{\mathrm{eff}}\widehat{\mathbf{\sigma }}\right)
D_{0}(\left\vert \mathbf{r-r}_{0}\right\vert ),  \label{D_mag}
\end{equation}%
where $g$ is the constant describing the non-magnetic part of the
interaction (for $g>0$ the potential is repulsive), $J$ is the constant of
exchange interaction, $\mathbf{\mu }_{\mathrm{eff}}=\mu _{\mathrm{eff}}(\sin
\alpha ,0,\cos \alpha )$ is the magnetic moment of the cluster, $\widehat{%
\mathbf{\sigma }}=\left( \widehat{\sigma }_{x},\widehat{\sigma }_{y},%
\widehat{\sigma }_{z}\right) $ with $\widehat{\sigma }_{\mu }$ the Pauli
matrixes, and $\widehat{I}$ is the unit matrix. The function $D(\mathbf{r})$
satisfies the condition (\ref{intD=1}). In the case of spin flip scattering
the spinor electron wave functions satisfy the Schr\"{o}dinger equation (\ref%
{Schrod}), in which the scattering potential must be replaced by the matrix $%
\widehat{D}(\mathbf{r})$ (\ref{D_mag}). Under the assumptions that the
potential $\widehat{D}(\mathbf{r})$ and the transparency of the tunnel
barrier in the contact plane are small the two-component wave function can
be found by the method described in Sec.~II.

The difference in absolute values of the wave vectors $\mathbf{k}_{\sigma }$
for spin-up and spin-down electrons (for the same energy $\varepsilon $),
which move towards the contact from the ferromagnetic bank, 
\begin{equation}
k_{\uparrow \downarrow }=\frac{1}{\hbar }\sqrt{2m^{\ast }\left( \varepsilon
\mp 4\pi g_{e}\mu _{\mathrm{B}}M\right) },  \label{k_sigm}
\end{equation}%
results in different amplitudes $t_{\sigma }=t(\mathbf{k}_{\sigma })$ (see
Eq.~(\ref{t(k)})) of the electron waves injected into the non-magnetic metal
for different directions of the spin ($g_{e}$ is the electron g-factor). The
total effective polarization $P_{\mathrm{eff}}$ of the current depends on
the difference between the probabilities of tunnelling for different $\sigma 
$, 
\begin{equation}
P_{\mathrm{eff}}(\varepsilon )=\frac{\left\vert t_{\uparrow }\right\vert
^{2}-\left\vert t_{\downarrow }\right\vert ^{2}}{\left\vert t_{\uparrow
}\right\vert ^{2}+\left\vert t_{\downarrow }\right\vert ^{2}}.  \label{P}
\end{equation}

\begin{figure}[tbp]
\includegraphics[width=8cm,angle=0]{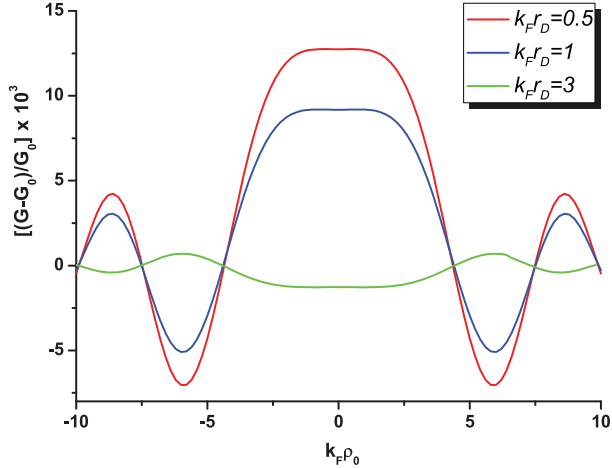}
\caption{Dependence of the the oscillatory part of the conductance on the
tip position on the metal surface for a subsurface magnetic cluster with
different cluster diameters. The $\protect\rho _{0}$-coordinate is measured
from the point $\protect\rho _{0}=0$ at which the contact is situated
directly above the cluster; $r_{0}=(0,0,10)/k_{\mathrm{F}};$ $\widetilde{g}%
=0.5;$ $\widetilde{J}=\frac{m^{\ast }k_{\mathrm{F}}}{\protect\mu _{\mathrm{B}%
}\hbar ^{2}}J\protect\mu _{\mathrm{eff}}=2.5;$ $P_{eff}=0.4$; $\protect%
\alpha =0$ \protect\cite{Avotina09}. }
\label{Fig52}
\end{figure}

The conductance $G$ of the contact at $T=0$ and $eV\ll \varepsilon _{\mathrm{%
F}}$ is given by \cite{Avotina09} 
\begin{equation}
G=\frac{I}{V}=G_{0}\left[ 1+\frac{6m^{\ast }k_{\mathrm{F}}}{\pi \hbar ^{2}}%
\left( g+\frac{1}{2\mu _{\mathrm{B}}}P_{\mathrm{eff}}\left( \varepsilon _{%
\mathrm{F}}\right) J\cos \alpha \right) W(\mathbf{r}_{0})\right]
_{\varepsilon =\varepsilon _{\mathrm{F}}},  \label{G(alpha)}
\end{equation}%
where $G_{0}$ is the conductance of the contact in absence of the cluster%
\begin{equation}
G_{0}=\left( k_{\mathrm{F}\uparrow }^{2}+k_{\mathrm{F}\downarrow
}^{2}\right) \frac{e^{2}\hbar ^{3}\left( k_{\mathrm{F}}a\right) ^{4}}{72\pi
\left( m^{\ast }U_{0}\right) ^{2}},  \label{G_0}
\end{equation}%
$k_{\mathrm{F}\sigma }$ is the absolute value of the Fermi wave vector in
the magnetic metal for spin direction $\sigma $ (see Eq.~(\ref{k_sigm})),
and 
\begin{equation}
W(\mathbf{r}_{0})=\int d\mathbf{r}^{\prime }D_{0}(\left\vert \mathbf{r}%
^{\prime }\mathbf{-r}_{0}\right\vert )\left( \frac{z^{\prime }}{r^{\prime }}%
\right) ^{2}w(kr^{\prime }).  \label{W}
\end{equation}%
The function $w$ is defined by Eq.~(\ref{w}). When the radius of action $%
r_{D}$ of the function $D_{0}(\left\vert \mathbf{r-r}_{0}\right\vert )$ is
much smaller than the distance between the contact and the center of the
cluster, $r_{0},$ $W(\mathbf{r}_{0})$ is an oscillatory function of $kr_{0}$
for $kr_{D}\geq 1,$ as for point defect with $kr_{D}\ll 1$ (see, Eq.~(\ref%
{GVa->0}) at $V=0$), but the oscillation amplitude is reduced as a result of
superposition of waves scattered by different points of the cluster. The
integral $W(\mathbf{r}_{0})$ (\ref{W}) can be calculated asymptotically for $%
r_{0}\gg r_{D},$ $kr_{0}\gg 1,$ and $kr_{D}\gtrsim 1.$ For a homogeneous
spherical potential $D_{0}(\left\vert \mathbf{r}\right\vert
)=V_{D}^{-1}\Theta (r_{D}-r)$ ($V_{D}$ is the cluster volume) the function $%
W(\mathbf{r}_{0})$ takes the form 
\begin{equation}
W(\mathbf{r}_{0})\simeq 3\left( \frac{z_{0}}{r_{0}}\right) ^{2}\frac{\sin
2kr_{0}}{\left( 2kr_{0}\right) ^{2}}\frac{j_{1}(kd)}{kd},  \label{Wasym}
\end{equation}%
where $d=2r_{D}$ is the cluster diameter. The last factor in Eq.~(\ref{Wasym}%
) describes the quantum size effect related with electron reflections by the
cluster boundary. Such oscillations may exist, if the cluster boundary is
sharp on the scale of the electron wave length. Fig.~\ref{Fig52} shows the
dependence of the amplitude of the conductance oscillations on the cluster
diameter. It demonstrates that a $\pi $-phase shift may occur resulting from
interference of electron waves over a distance of the cluster diameter. 

In Eq.~(\ref{G(alpha)}) the term proportional to $P_{\mathrm{eff}}$ takes
into account the difference in the probabilities of scattering of electrons
with different $\sigma $ by the localized magnetic moment $\mathbf{\mu }_{%
\mathrm{eff}}.$ It depends on the angle $\alpha $ between the tip
magnetization and $\mathbf{\mu }_{\mathrm{eff}}$, as $\cos \alpha .$ The
same dependence was first predicted for a tunnel junction between
ferromagnets for which the magnetization vectors are misaligned by an angle $%
\alpha $ \cite{Slonczewski}, and this was observed in SP-STM experiments 
\cite{Bode}.

Note that once the spin-polarized current-induced torque pulls the magnetic
moment away from alignment with $H$, the cluster moment will start
precessing around the field axis. The Larmor frequency is defined by the
magnetic field due to combining the external field $H$ and the effective
magnetic field produced by the polarized current. The precession of the
cluster magnetic moment gives rise to a time modulation of the SP-STM
current as for clusters on a sample surface \cite{Manoharan,Durkan}.

\section{Magneto-quantum oscillations}

\subsection{Conductance oscillations in perpendicular magnetic field}

In a strong magnetic field the STM conductance exhibits characteristic
oscillations in magnetic field, which are attributed to Landau quantization.
This effect has been observed in Ref.~\cite{Morgenstern} and the energy
dependence of the effective electron mass was determined. An influence of
the magnetic field on the interference pattern, which is produced by two
adatoms, in the STM conductance has been investigated theoretically \cite%
{Cano} and horizontal stripes related to the Aharonov-Bohm effect were
predicted.

In Sec.~III it is demonstrated that the dependence $G(\mathbf{r}_{0},V)$
undergoes oscillations in $r_{0}$ and $eV$ resulting from the variation of
the phase shift between transmitted and scattered electron waves. Here we
discuss another way to control the phase shift between the interfering
waves: an applied external magnetic field $\mathbf{H}$ produces oscillations
of the conductance as a function of $H\mathbf{.}$

Let us consider the contact described in Sec.~II, now placed in a magnetic
field directed along the contact axis, $\mathbf{H}=(0,0,H).$ Figure~\ref%
{traject_mag_field} shows schematically the trajectories of the electrons
that are injected into the metal and interact with the defect. 
\begin{figure}[tbp]
\includegraphics[width=8cm,angle=0]{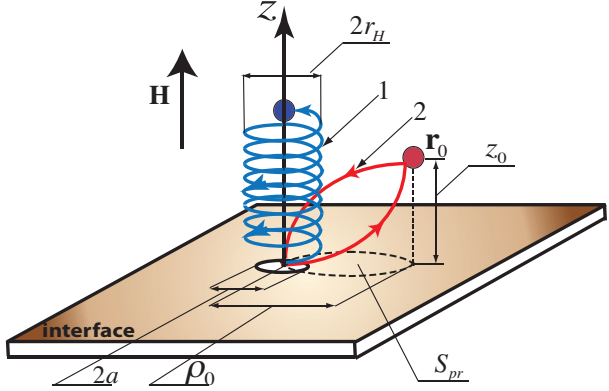}
\caption{Schematic representation of the electron trajectories in a vicinity
of a point contact in an external magnetic field oriented along the contact
axis.}
\label{traject_mag_field}
\end{figure}

In what follows the Schr\"{o}dinger equation is solved along the same lines
as in Sec.~II, and as zeroth approximation we use the well-known wave
function for an electron in a homogeneous magnetic field. In Ref.~\cite%
{Avotina07} the dependence of the STM conductance on magnetic field has been
obtained under the assumptions that the contact diameter $a$ is much smaller
than the magnetic quantum length, $a_{H}=\sqrt{\hbar /m^{\ast }\Omega },$
the radius of the electron trajectory, $r_{H}=\hbar k_{\mathrm{F}}/m^{\ast
}\Omega ,$ is much smaller than the mean free path of the electrons, $l\gg
r_{0}$, and the separation between the magnetic quantum levels, the Landau
levels, $\hbar \Omega $ is larger than the temperature $k_{\mathrm{B}}T$ , ($%
\Omega =eH/m^{\ast }c$ is the Larmor frequency). Although these conditions
restrict the possibilities for observing the oscillations severely, all
conditions can be realized, e.g., in single crystals of semimetals (Bi, Sb
and their ordered alloys) where the electron mean free path can be up to
millimeters and the Fermi wave length $\lambda _{\mathrm{F}}\sim 10^{-8}$m.
Under condition of the inequalities listed the dependence of the conductance
of the tunnel point contact on $H$ is given by \cite{Avotina07} 
\begin{equation}
G(H)=G_{c}\left( H\right) \left[ 1+\frac{gm^{\ast }}{2\pi ^{3}\left( N_{%
\mathrm{F}\uparrow }+N_{\mathrm{F}\downarrow }\right) \hbar ^{2}a_{H}^{4}}%
\sum\limits_{\sigma }\left( \func{Im}\sum\limits_{n=0}^{n_{\max }}\chi
_{\sigma }(n,\mathbf{r}_{0})\right) \left( \func{Re}\sum\limits_{n^{\prime
}=0}^{\infty }\chi _{\sigma }(n^{\prime },\mathbf{r}_{0})\right) \right] .
\label{G(H)}
\end{equation}%
Here 
\begin{equation}
\chi _{\sigma }(n,\mathbf{r}_{0})=\exp \left( -\frac{\xi _{0}}{2}\right)
L_{n}(\xi _{0})\exp \left( \frac{i}{\hbar }z_{0}\sqrt{2m^{\ast }\left(
\varepsilon _{\mathrm{F}}+\sigma \mu _{\mathrm{B}}H-\varepsilon _{n}\right) }%
\right) ,  \label{hi}
\end{equation}%
$\xi _{{0}}=\rho _{0}^{2}/2a_{H}^{2}$, and $L_{n}(\xi )$ are Laguerre
polynomials, $\varepsilon _{n}=\hbar \Omega \left( n+{\frac{1}{2}}\right) $, 
$\sigma =\pm 1\ $is the spin index, $N_{\mathrm{F}\sigma }$ is the number of
electron states for one spin direction per unit volume at the Fermi energy,%
\begin{equation}
N_{\mathrm{F}\sigma }=\frac{2\left\vert e\right\vert H}{\left( 2\pi \hbar
\right) ^{2}c}\sum\limits_{n=0}^{n_{\max }}\sqrt{2m^{\ast }\left(
\varepsilon _{\mathrm{F}}+\sigma \mu _{\mathrm{B}}H-\varepsilon _{n}\right) }%
,  \label{N_F}
\end{equation}%
$n_{\max }=\left[ \frac{\varepsilon _{\mathrm{F}}}{\hbar \Omega }\right] $
is the maximum value of the quantum number $n$ for which $\varepsilon
_{n}<\varepsilon _{\mathrm{F}},$ and $\left[ x\right] $ is the integer part
of the number $x$, $G_{c}$ is the conductance in absence of a defect, 
\begin{equation}
G_{c}(H)=\left( \pi \hbar \right) ^{3}\left( \frac{ea^{2}\left( N_{\mathrm{F}%
\uparrow }+N_{\mathrm{F}\downarrow }\right) }{m^{\ast }U_{0}}\right) ^{2}.
\label{G_c(H)}
\end{equation}%
The conductance (\ref{G_c(H)}) undergoes oscillations having the periodicity
of the de Haas-van Alphen effect that originates from the step-wise
dependence of the number of electron states $N_{\mathrm{F}\sigma }$ (\ref%
{N_F}) on the magnetic field. At $n_{\max }(\varepsilon _{\mathrm{F}})\gg 1,$
$\mu _{\mathrm{B}}H/\varepsilon _{\mathrm{F}}\ll 1$ (semiclassical
approximation), Eq.(\ref{G_c(H)}) can be expanded in the small parameter $%
\hbar \Omega /\varepsilon _{\mathrm{F}}$ 
\begin{equation}
G_{c}(H)\simeq G_{0}\left[ 1+\frac{9}{2}\left( \frac{\hbar \Omega }{%
\varepsilon _{\mathrm{F}}}\right) ^{3/2}\sum\limits_{s=1}^{\infty }\frac{%
\left( -1\right) ^{s}}{\left( 2s\right) ^{3/2}}\sin \left( 2\pi s\frac{%
\varepsilon _{\mathrm{F}}}{\hbar \Omega }-\frac{\pi }{4}\right) \right] ,
\end{equation}%
where $G_{0}$ is the conductance of the contact at $H=0$ (see Eq.~(\ref{G_0}%
)).

The oscillatory part of the conductance, $\Delta G(H)=G(H)-G_{c}(H),$ which
results from the electron scattering on the defect, is plotted in Fig.~\ref%
{fig_G(H)} for a defect placed at $(\rho ,z)=(50,30)/k_{\mathrm{F}}$. Figure~%
\ref{fig_G(Hr0)} illustrates the dependence of the conductance (\ref{G(H)})
on $\rho _{0}$ coordinate of the defects for different $H.$ The beating of
the oscillation amplitude due to the difference of electron energies for
different spin is seen at higher magnetic field. 
\begin{figure}[tbp]
\includegraphics[width=8cm]{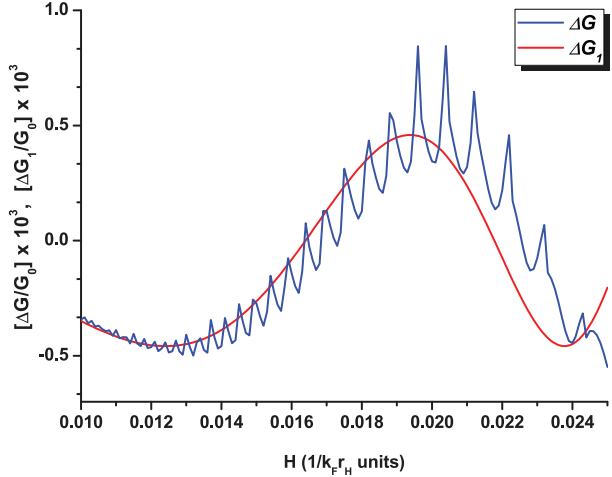}
\caption{Oscillatory part of the conductance of a tunneling point contact
with a single defect placed at $k_{\mathrm{F}}\protect\rho _{0}=50,$ $k_{%
\mathrm{F}}z_{0}=30.$ The full curve is a plot for Eq.~(\protect\ref{G(H)}),
while the dashed curve shows the component $\Delta G_{1}$ for the
semiclassical approximation, Eq.~(\protect\ref{dG1}). The field scale is
given in units $1/k_{\mathrm{F}}r_{H}$; $\widetilde{g}=0.5$. }
\label{fig_G(H)}
\end{figure}
\begin{figure}[tbp]
\includegraphics[width=8cm]{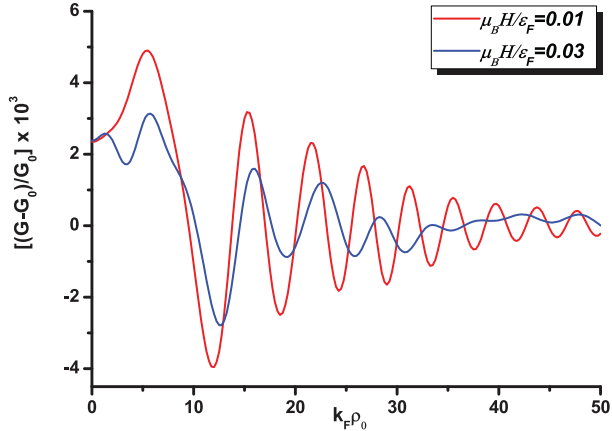}
\caption{Dependence of the oscillatory part of the STM conductance on the
tip position for different values of magnetic field, $z_{0}=30/k_{\mathrm{F}%
},$ $\widetilde{g}=0.5.$ }
\label{fig_G(Hr0)}
\end{figure}

The dependence plotted in Fig.~\ref{fig_G(H)}, $G\left( H\right),$ (\ref%
{G(H)}), contains oscillations with different periods. The semiclassical
asymptotes at $\hbar \Omega \ll \varepsilon _{\mathrm{F}}$ of the expression
for the conductance Eq.~(\ref{G(H)}) allows us to explain the physical
origin of these oscillations. By using the Poisson summation formula in Eq.~(%
\ref{G(H)}) the part of the conductance $\Delta G(H)$ related with the
scattering by the defect can be written as a sum of two terms 
\begin{equation}
\Delta G(H)=\Delta G_{1}+\Delta G_{2},  \label{sum}
\end{equation}%
each of them describing conductance oscillations with different periods,
which are discussed in more detail below. \qquad

\subsection{Effect of flux quantization through the trajectory of the
scattered electrons}

The first term $\Delta G_{1}(H,\mathbf{r}_{0})$ in Eq.~(\ref{sum}) describes
the long-period oscillations 
\begin{equation}
\Delta G_{1}(H,\mathbf{r}_{0})=-G_{0}\widetilde{g}\frac{z_{0}^{2}}{k_{%
\mathrm{F}}^{2}r_{0}^{4}}\sin \left( 2k_{\mathrm{F}}r_{0}-2\pi \frac{\Phi }{%
\Phi _{0}}\right) .  \label{dG1}
\end{equation}%
where $\Phi _{0}=2\pi \hbar c/e$ is the flux quantum. The flux, $\Phi =HS_{%
\mathrm{pr}},$ is given by the field lines penetrating the areas of the
projections $S_{\mathrm{pr}}$ on the plane $z=0$ of the trajectories of the
electrons moving from the contact to the defect and back (see, the
trajectory 2 in Fig.~\ref{traject_mag_field}). Such trajectories consist of
two arcs, and there are a lot of trajectories with different $S_{\mathrm{pr}%
}.$ As was shown in Refs.~\cite{Avotina07} among these trajectories the
signal is dominated by the one that has minimal area given by $S_{\mathrm{pr}%
}=2S_{\mathrm{seg}}$. Here $S_{\mathrm{seg}}=r^{2}\left( \theta -\sin
2\theta \right) $ is the area of the segment formed by the chord of length $%
\rho _{0}$ and the arc of radius $r=r_{H}\sin \theta $, with $\theta $ the
angle between the vector $\mathbf{r}_{0}$ and the $z$-axis, $\sin \theta
=\rho _{0}/r_{0}.$ Therefore, the oscillation $\Delta G_{1}$ disappears when
the defect sits on the contact axis, $\rho _{0}=0.$ Obviously, the origin of
these oscillations lies in the curvature of the electron trajectories in a
magnetic field. As we can see from Eq.~(\ref{dG1}) the oscillations in the
conductance $\Delta G_{1}$ have a nature similar to the Aharonov-Bohm effect
(the conductance undergoes oscillations with a period $\Phi /\Phi _{0}$) and
are related to the quantization of the magnetic flux through the area
enclosed by the electron trajectory. For illustration of this fact in Fig.~%
\ref{fig_G(H)} the full expression for the oscillatory part $\Delta G(H)$ of
the conductance (the second term in Eq.~(\ref{G(H)})) is compared with the
semiclassical approximation $\Delta G_{1}(H,\rho _{0},z_{0})$, Eq~(\ref{dG1}%
).

For the observation of the Aharonov-Bohm-type oscillations the position $%
\rho _{0}$ of the defect in the plane parallel to the interface must be
smaller then $r_{H}$, i.e. the defect must be situated inside the `tube' of
electron trajectories passing through the contact. At the same time the
inequality $\rho _{0}>a_{H}$ must hold in order that a magnetic flux quantum 
$\Phi _{0}$ is enclosed by the area of the closed trajectory.

\subsection{Effect of longitudinal focusing of electrons on the defect by
the magnetic field}

The short-period oscillations originate from the effect of the electron
being focused by the magnetic field, and is described by the term $\Delta
G_{2}(H,\mathbf{r}_{0})$ in Eq.~(\ref{sum}). At $\rho _{0}=0$ this term can
be written as 
\begin{equation}
\Delta G_{2}(H,z_{0})\simeq \frac{1}{16}G_{0}\widetilde{g}\left( \frac{\hbar
\Omega }{2\varepsilon _{\mathrm{F}}}\right) ^{3/2}\sum\limits_{s=\left[
z_{0}/2\pi r_{H}\right] }^{\infty }\frac{\left( -1\right) ^{s}}{s^{3/2}}\cos
\left( k_{\mathrm{F}}r_{0}+2\pi s\frac{\varepsilon _{\mathrm{F}}}{\hbar
\Omega }+\frac{z_{0}^{2}}{4\pi sa_{H}^{2}}\right) .  \label{dG2}
\end{equation}%
In the absence of a magnetic field only those electrons that are scattered
off the defect in the direction directly opposite to the incoming electrons
can come back to the point-contact. When $H\neq 0$ the electrons move along
a spiral trajectory and may come back to the contact after scattering under
a finite angle to the initial direction (the trajectory 1 in Fig.8). For
example, if the defect is placed on the contact axis an electron moving from
the contact with a wave vector $k_{z}=k_{\mathrm{F}}$ along the magnetic
field returns to the contact when the $z$-component of the momentum $%
k_{zs}=z_{0}m^{\ast }\Omega /2\pi s\hbar $, for integer $s$. For these
orbits the time of the motion over a distance $z_{0}$ in the $z$ direction
is a multiple of the cyclotron period $T_{H}=2\pi /\Omega $. Thus, after $s$
revolutions the electron returns to the contact axis at the point $z=0.$ The
phase which the electron acquires along the spiral trajectory is composed of
two parts, $\Delta \phi =\Delta \phi _{1}+\Delta \phi _{2}.$ The first, $%
\Delta \phi _{1}=k_{zs}z_{0}$ is the `geometric' phase accumulated by an
electron with wave vector $k_{zs}$ over the distance $z_{0}$. The second, $%
\Delta \phi _{2}=\pi s(eHr_{s}^{2}/c\hbar )$ is the phase acquired during $s$
rotations in the field $H,$ where $r_{s}=\hbar c\sqrt{k_{\mathrm{F}%
}^{2}-k_{zs}^{2}}/eH$ is the radius of the spiral trajectory. Substituting $%
k_{zs}$ and $r_{s}$ in the equation for $\Delta \phi $ we find%
\begin{equation}
\Delta \phi =2\pi s\varepsilon _{\mathrm{F}}/\hbar \Omega +z_{0}^{2}/4\pi
sa_{H}^{2}.
\end{equation}%
This is just the phase shift that defines the period of oscillation in the
contribution $\Delta G_{2}$ (\ref{dG2}) to the conductance. It describes a
trajectory which is straight for the part from the contact to the defect and
spirals back to the contact by $s$ windings as it is shown in Fig.8. There
are trajectories consisting of helices in the forward and reverse paths,
with $s$ and $s^{\prime }$ coils, respectively. However, the contribution of
these trajectories to the conductance is smaller than $\Delta G_{2}$ (\ref%
{dG2}) by a factor $\sim 1/\left( k_{\mathrm{F}}a_{H}\right) \ll 1.$ Note
that, although the amplitude of the oscillation $\Delta G_{2}$ (\ref{dG2})
is smaller by a factor $\hbar \Omega /\varepsilon _{\mathrm{F}}$ than the
amplitude of the contribution $\Delta G_{1}$ (\ref{dG1}), the first depends
on the depth of the defect as $z_{0}^{-3/2}$ while $\Delta G_{1}\sim
z_{0}^{-2}.$ The slower decreasing of the amplitude for $\Delta G_{2}$ is
explained by the effect of focusing of the electrons in the magnetic field.
The predicted oscillations, Eq.~(\ref{dG2}), are not periodic in $H$ nor in $%
1/H$. Their typical period can be estimated as the difference $\Delta H$
between two nearest-neighbor maxima 
\begin{equation}
\left( \frac{\Delta H}{H}\right) \simeq \frac{\hbar \Omega }{\varepsilon _{%
\mathrm{F}}}\left( 1-\left( \frac{z_{0}}{2\pi k_{\mathrm{F}}a_{H}^{2}}%
\right) ^{2}\right) ^{-1}.  \label{sp}
\end{equation}%
The period (\ref{sp}) depends on the position of the defect. It is larger
than the period of de Haas-van Alphen oscillation, $\left( \Delta H/H\right)
_{dHvA}\simeq \hbar \Omega /\varepsilon _{\mathrm{F}}.$ Both of these
periods are of the same order of magnitude.

\section{Nonmagnetic defect in a superconductor}

In this section we present the results of a theoretical investigation of the
conductance, $G_{ns}$, of a normal metal - superconductor (NS) point contact
(with radius $a<\lambda _{F}$ ) in the tunnelling limit and discuss the
quantum interference effects originating from the scattering of
quasiparticles by a point-like nonmagnetic defect \cite{Avotina08ns}. The
model is described in Sec. II and illustrated in Fig.~\ref{Contact},
modified with having the half-space $z>0$ occupied by a ($s$-wave)
superconductor. At zero temperature a tunnel current flows through the
contact for an applied bias $eV$ larger than the energy gap of the
superconductor $\Delta _{0}$. In order to evaluate the total current through
the contact, $I\left( V\right) $ , the current density $\mathbf{j}_{\mathbf{k%
}}\left( \mathbf{r}\right) $ of quasiparticles with momentum $\mathbf{k}$ at 
$z>0,$ formed by electrons transmitted through the contact must be found.
The current density $\mathbf{j}_{\mathbf{k}}\left( \mathbf{r}\right) $ is
expressed in terms of the coefficients $u_{\mathbf{k}}\left( \mathbf{r}%
\right) $ and $v_{\mathbf{k}}\left( \mathbf{r}\right) $ of the canonical
Bogoliubov transformation \cite{Hard}. The functions $u_{\mathbf{k}}\left( 
\mathbf{r}\right) $ and $v_{\mathbf{k}}\left( \mathbf{r}\right) $ satisfy
the Bogoliubov-de Gennes (BdG) equations \cite{Gennes}, which must be
supplemented with a self-consistency condition for the order parameter $%
\Delta \left( \mathbf{r}\right) $, and boundary conditions which connect $u_{%
\mathbf{k}}$ and $v_{\mathbf{k}}$ in the normal metal to those in the
superconductor at the contact. For a tunnel contact one can neglect Andreev
reflections, because these lead to corrections to the conductance
proportional to $\left\vert t\right\vert ^{4}$ \cite{BTK}; the functions $u_{%
\mathbf{k}}$ and $v_{\mathbf{k}}$ satisfy the same boundary conditions (\ref%
{equal2}), (\ref{jump2}) as the wave function for a contact between normal
metals.

It is obvious that the method described in Sec.~II and Sec.~III can be
generalized to NS contacts. As a first step the Bogoliubov-de Gennes (BdG)
equations must be solved in linear approximation in the transmission
amplitude $t$ in the absence of the defect, $D=0,$ after which the
corrections due to the scattering by the defect can be found. In Ref.~\cite%
{Avotina08ns} an analytical solution for the BdG equations was found with
the approximation of a homogeneous order parameter $\Delta \left( \mathbf{r}%
\right) =\Delta _{0}\Theta \left( z\right) .$

At small applied bias $eV\ll \hbar \omega _{\mathrm{D}}\ll \varepsilon _{%
\mathrm{F}}$ ($\omega _{\mathrm{D}}$ is the Debye frequency), and in linear
approximation in the electron-defect interaction constant $g$ the
conductance $G_{ns}$ \ of a NS tunnel point contact can be presented as the
sum of two terms, 
\begin{equation}
G\left( V,r_{0}\right) =G_{0ns}\left( V\right) +\Delta G_{osc}\left(
V,r_{0}\right) ,\quad eV>\Delta _{0}.  \label{G}
\end{equation}%
The first term, $G_{0ns}\left( V\right) $, in Eq.~(\ref{G}) is the
conductance of the NS tunnel point contact in the absence of the defect 
\begin{equation}
G_{0ns}\left( V\right) =G_{0}N_{s}\left( eV\right) ,  \label{G0_ns}
\end{equation}%
where $G_{0}$ is the conductance of a contact between normal metals (\ref{G0}
), which is multiplied by the normalized density of states of the
superconductor $N_{s}\left( E\right) =E/\sqrt{E^{2}-\Delta _{0}^{2}}$ at $%
E=eV$. Although such result is not unexpected and has been confirmed by
experiment \cite{Pan}, for a contact of radius $k_{\mathrm{\ F}}a<1$ it was
not obvious and it is first obtained in Ref.~\cite{Avotina08ns}. The second
term describes the oscillatory dependence of the conductance with the
distance between the contact and the defect. If the defect is situated in
the superconductor $\left( z_{0}<0\right) $ 
\begin{equation}
\Delta G_{osc}\left( V,r_{0}\right) =-G_{0ns}\left( V\right) \widetilde{g}%
\left( \frac{z_{0}}{r_{0}}\right) ^{2}\sum\limits_{\alpha =\pm }\psi
_{\alpha }\left( eV\right) w\left( k_{\alpha }r_{0}\right) ,  \label{dG_S}
\end{equation}%
where%
\begin{equation}
\psi _{\pm }\left( eV\right) =\frac{1}{2}\left( 1\pm \frac{\sqrt{\left(
eV\right) ^{2}-\Delta _{0}^{2}}}{eV}\right) ,
\end{equation}%
\begin{equation}
k_{\pm }=\frac{\sqrt{2m^{\ast }}}{\hbar }\left[ \varepsilon _{\mathrm{F}}\pm 
\sqrt{\left( eV\right) ^{2}-\Delta _{0}^{2}}\right] ^{1/2},  \label{psi_k}
\end{equation}%
and the function $w\left( k_{\alpha }r_{0}\right) $ is given by Eq.~(\ref{w}%
). Equation~(\ref{dG_S}) is obtained by neglecting all small terms of the
order of $\Delta _{0}/\varepsilon _{\mathrm{F}}$ and $eV/\varepsilon _{%
\mathrm{F}}$. Note that we kept the second term in square brackets in the
formula for $k_{\pm }$ (Eq.~(\ref{psi_k})) because for large $r_{0},$ $(%
\sqrt{(eV)^{2}-\Delta _{0}^{2}}/\varepsilon _{\mathrm{F}})(k_{\mathrm{F}%
}r_{0})\simeq 1,$ the phase shift of the oscillations may be important.

\section{Conclusions}

Thus, we have reviewed some theoretical aspects of the possibility of
investigating subsurface defects by STM experiments. The theoretical results
show that the amplitude of the oscillations of the STM conductance resulting
from quantum interference of electron waves injected by the STM tip and
scattered by the defect remains sufficiently large ($\sim 10^{-3}G_{0}$),
even for defects located more than 10 atomic layers below the surface. For
example, in the STM experiments of Ref.~\cite{Stipe} signal-to-noise ratios
of $5\cdot 10^{-4}$ (at 1~nA, 400~Hz sample frequency) have been achieved.
Recently, the possibility of observing defects at such depths below the
surface has been demonstrated in experiment \cite{Weismann}.

The STM tip plays the role of "locator", which detects a defect below the
metal surface by using electron waves. The defect in turn produces
information about its (defect) characteristics, as well as showing
properties of the host metal by producing Friedel-like oscillations in the
STM conductance. The phase of the oscillations, $2k_{\mathrm{F}}r_{0},$ is
defined by the Fermi wave vector $k_{ \mathrm{F}}$ and the tip-defect
distance $r_{0}.$ One of the possibilities to determine the defect depth $%
z_{0}$ below surface is by changing the maximal value of the wave vector by
accelerating the electrons with an applied bias $eV$ \cite%
{Avotina05,Avotina09ltp25}. When the tip is situated above the defect, the
period of the oscillations in $G\left( V\right) $, $\Delta k_{\mathrm{F}%
}\left( eV\right) z_{0}=\pi $, uniquely defines $z_{0}.$ As the period of
the oscillations becomes longer for small $z_{0}$ the minimum detectable
depth will be determined by the maximum voltage that can be applied over the
junction. For example, 30 mV is sufficient for probing a quarter of a
conductance oscillation caused by a defect at 1~nm depth.

Another factor in setting the oscillation phase is the shape of the Fermi
surface (FS). As was shown, for an anisotropic FS $\varepsilon \left( 
\mathbf{k}\right) =\varepsilon _{\mathrm{F}}$ the phase and amplitude of
conductance oscillations depend on the characteristics of the FS in the
point for which the direction of the velocity $\mathbf{v}=\mathbf{v}_{0}$ is
parallel to the vector $\mathbf{r}_{0}$ directed from the STM tip to the
defect \cite{Avotina06}. Namely, the phase of the oscillations is defined by
the projection of $\mathbf{k}$ on the direction of $\mathbf{v}_{0},$ and the
oscillation amplitude depends on the curvature of the FS. Depending on the
geometry of the FS there can be several points with the same direction of
the velocity, or, if the FS has open parts, certain directions of the
velocity can be forbidden. It follows from the results above that curves of
constant phase $\mathbf{kv}_{0}r_{0}/\left\vert \mathbf{v}_{0}\right\vert $
(maxima and minima) in the interference pattern of the STM conductance show
the contours formed by projections of the vector $\mathbf{k}$ on the vector
normal to the FS. Although such contours reflect the main features of the FS
geometry, they cannot be considered as a direct imaging of the FS.

Electron scattering by subsurface magnetic defects in STM conductance
possesses some features distinct from the scattering by magnetic adatoms and
the form and sign of the Kondo anomaly due to a subsurface magnetic defect
depend on the depth \cite{Avotina08mi}. Near the Kondo resonance the
scattering phase shift $\delta _{0}$ tends to $\pi /2$, and including
multiple electron scattering events after reflections by the metal surface
becomes essential. This explains the appearance of harmonics in the
oscillatory part of the conductance, which have an additional phase shift $%
\Delta \phi =2\left( n-1\right) k_{\mathrm{F}}z_{0}+n\delta _{0},$ where $n$
is the number of electron reflections by the surface. The determination of
this phase shift near the Kondo resonance $\left( V\simeq V_{\mathrm{K}%
}\right) $ and far from it (where $\delta _{0}\ll 1,$) for the first $\left(
n=0\right) $ and second $\left( n=1\right) $ harmonics provides an
alternative way to find the depth of the defect $z_{0}.$

The possibilities of investigating magnetic defects are extended by
injecting a spin-polarized current. If the subsurface cluster possesses an
unscreened magnetic moment $\mathbf{\mu }_{\mathrm{eff}}$, the scattering
amplitudes of spin-up and spin-down electrons are different. This results in
a dependence of the oscillation amplitude on the angle between the vector $%
\mathbf{\mu }_{\mathrm{eff}}$ and the polarization direction of the STM
current - referred to as a magneto-orientational effect \cite{Avotina09}.

A strong magnetic field perpendicular to the metal surface changes the
interference pattern in the STM conductance fundamentally. As a result of
Zeeman splitting $\pm g_{e}\mu _{\mathrm{B}}H$ of the Landau energy levels
the interference patterns formed in the dependence $G\left( \mathbf{r}%
_{0}\right) $ by electrons with different spin directions do not coincide
because of different electron wave lengths for energies $\varepsilon _{%
\mathrm{F}}\pm g_{e}\mu _{\mathrm{B} }H.$ The superposition of the two
oscillatory parts may result in a beating of the total amplitude of the
oscillations. Along with the well-known quantum oscillations having the
periodicity in $H^{-1}$ of the de Haas - van Alphen effect in the STM
conductance, in the presence of a defect two new types of oscillations are
present. The first is related to flux quantization through the projection of
the electron trajectory on the surface plane. The second type of oscillation 
$G\left( H\right) $ is related to a focusing effect of the magnetic field.
As in Sharvin's two-point contact experiments, in which electrons were
focused on a collector by a magnetic field directed along the line
connecting the contacts (geometry of longitudinal electron focusing) \cite%
{Sharvin}, the magnetic field can periodically focus the electrons injected
by the tip onto the defect. This results in periodic increasing or
decreasing of the part of the conductance related to the scattering by the
defect \cite{Avotina07}.

If the electrons tunnel from a normal-metal STM tip into a superconductor
the wave incident on the contact is transformed into a superposition of
`electron-like' and `hole-like' quasiparticles. In the case of a location of
the defect in the superconductor quantum interference takes place between
the partial wave that is transmitted and the one that is scattered by the
defect, for both types of quasiparticles independently (Eq.~(\ref{dG_S})).
Although the difference between wave vectors $k^{\left( \pm \right) }\left(
eV\right) $ of `electrons' and `holes' is small the shift $\left( k^{\left(
+\right) }-k^{\left( -\right) }\right) r_{0}$ between the two oscillations
should be observable \cite{Avotina08ns}.

\end{document}